\documentclass[final]{agujournal2019}
\usepackage{url} 
\usepackage{lineno}
\usepackage{amsmath,amsfonts,amssymb}
\usepackage[inline]{trackchanges} 
\usepackage{soul}
\usepackage[normalem]{ulem}
\usepackage{cancel}

\newcommand*\patchAmsMathEnvironmentForLineno[1]{
  \expandafter\let\csname old#1\expandafter\endcsname\csname #1\endcsname
  \expandafter\let\csname oldend#1\expandafter\endcsname\csname end#1\endcsname
  \renewenvironment{#1}
  {\linenomath\csname old#1\endcsname}
  {\csname oldend#1\endcsname\endlinenomath}}
  \newcommand*\patchBothAmsMathEnvironmentsForLineno[1]{
  \patchAmsMathEnvironmentForLineno{#1}
  \patchAmsMathEnvironmentForLineno{#1*}}
  \AtBeginDocument{
  \patchBothAmsMathEnvironmentsForLineno{equation}
  \patchBothAmsMathEnvironmentsForLineno{align}
  \patchBothAmsMathEnvironmentsForLineno{flalign}
  \patchBothAmsMathEnvironmentsForLineno{alignat}
  \patchBothAmsMathEnvironmentsForLineno{gather}
  \patchBothAmsMathEnvironmentsForLineno{multline}
}

\newcommand{\mycomment}[1]{}
\draftfalse

\journalname{Journal of Advances in Modeling Earth Systems (JAMES)}

\begin{document}

%
%


\title{A stable implementation of a data-driven scale-aware mesoscale parameterization}

%
%




\authors{Pavel Perezhogin\affil{1}, Cheng Zhang\affil{2}, Alistair Adcroft\affil{2}, Carlos Fernandez-Granda\affil{1,3}, Laure Zanna\affil{1}}


\affiliation{1}{Courant Institute of Mathematical Sciences, New York University, New York, NY, USA}
\affiliation{2}{Program in Atmospheric and Oceanic Sciences, Princeton University, Princeton, NJ 08542, USA}
\affiliation{3}{Center for Data Science, New York University, New York, NY, USA}




\correspondingauthor{P. Perezhogin}{pp2681@nyu.edu}




\begin{keypoints}
\item A data-driven mesoscale eddy parameterization is implemented and evaluated in different configurations of the GFDL MOM6 ocean model
\item We introduce filtering schemes to reduce the generation of grid-scale noise and enhance the large-scale backscatter

\item The subgrid parameterization improves the representation of the energy distributions and the climatological mean state
\end{keypoints}

%
%

%
%


\begin{abstract}
Ocean mesoscale eddies are often poorly represented in climate models, and therefore, their effects on the large scale circulation must be parameterized. Traditional parameterizations, which represent the bulk effect of the unresolved eddies, can be improved with new subgrid models learned directly from data. \citeA{zanna2020data} (ZB20) applied an equation-discovery algorithm to reveal an interpretable expression parameterizing the subgrid momentum fluxes by mesoscale eddies through the components of the velocity-gradient tensor. In this work, we implement the ZB20 parameterization into the primitive-equation GFDL MOM6 ocean model and test it in two idealized configurations with significantly different dynamical regimes and topography. The original parameterization was found to generate excessive numerical noise near the grid scale. 
We propose two filtering approaches to avoid the numerical issues and additionally enhance the strength of large-scale energy backscatter.  The filtered ZB20 parameterizations led to improved climatological mean state and energy distributions, compared to the current state-of-the-art energy backscatter parameterizations. The filtered ZB20 parameterizations are scale-aware and, consequently, can be used with a single value of the non-dimensional scaling coefficient for a range of resolutions. The successful application of the filtered ZB20 parameterizations to parameterize mesoscale eddies in two idealized configurations offers a promising opportunity to reduce long-standing biases in global ocean simulations in future studies. 
\end{abstract}

\section*{Plain Language Summary}
This research focuses on improving the accuracy of ocean models by addressing the challenges of representing the mesoscale eddies on coarse grids. These eddies play a crucial role in the Earth's climate system, but traditional climate models struggle to capture their effects. Here, we implemented a new data-driven parameterization simulating the physics of the mesoscale eddies into the state-of-the-art ocean model. The parameterization is interpretable and captures key physical processes related to the mesoscale eddies known as energy backscatter. We tested this parameterization in two idealized ocean scenarios and found that it significantly improves the biases in the representation of the mean state and energetics. We propose new filtering schemes which improve the physical and numerical properties of the parameterization. Accurate representation of the mesoscale eddies by the present scheme has the potential to resolve long-standing biases present in global ocean models and thus allow for more reliable climate simulations.

\section{Introduction}
Ocean mesoscale eddies emerge on the spatial scale of the Rossby deformation radius \cite{salmon1980baroclinic, vallis2017atmospheric}, which varies in the global ocean from order 100 km near the equator to 10 km near the poles \cite{chelton1998geographical}. Mesoscale eddies dominate the ocean kinetic energy (KE) reservoir and are critical for the lateral and vertical transport of tracers \cite{redi1982oceanic, ferrari2009ocean, uchida2017seasonality}. The momentum and buoyancy fluxes produced by the mesoscale eddies are crucial in strengthening the mean flow via upgradient fluxes \cite{greatbatch2010transport}, setting the stratification \cite{gent1990isopycnal}, and closing the global ocean energy budget \cite{jansen2019toward, bachman2019gm, loose2023diagnosing}.

To adequately simulate mesoscale eddies, several grid spacings per deformation radius are required \cite{hallberg2013using}. Such eddy-resolving simulations remain unfeasible for the foreseeable future in global coupled climate models \cite{hewitt2020resolving, christensen2022parametrization}. Therefore, the systematic effect of unresolved mesoscale eddies must be \textit{parameterized} to reduce the biases in the climatology, variability, and response to forcing of the ocean and climate. Traditional parameterizations mimic the bulk (that is, the mean) effect of the mesoscale eddies on the resolved flow and are often based on energetic considerations \cite<e.g.,>[]{jansen2015energy}. For example, the parameterization of \citeA{gent1990isopycnal} and its extensions \cite{marshall2012framework, mak2018implementation, mak2022acute, mak2023scale} reduces the potential energy by flattening the isopycnals. Another popular parameterization is the \textit{KE backscatter} \cite{frederiksen1997eddy, frederiksen1999subgrid, thuburn2014cascades, jansen2014parameterizing, grooms2015numerical, berloff2018dynamically, bachman2019gm, jansen2019toward, perezhogin2020testing, juricke2020ocean, storto2021new}, which returns energy from the subgrid scales to the larger scales.   Several of these parameterizations attempt to correct the dissipation associated with eddy viscosity models by returning the KE to the resolved flow. Other parameterizations represent the net inverse energy cascade from subgrid to resolved scales. 


Recently, machine-learning methods based on neural networks 
have been proposed to predict the eddy fluxes directly from data as an alternative to the traditional bulk modeling \cite{krasnopolsky2010accurate, rasp2018deep, yuval2020stable, guan2022stable, frezat2022posteriori, sane2023, perezhogin2023generative, yuval2023neural, shamekh2023implicit, zampieri2023machine, gregory2023deep}. Neural networks capture the bulk effect of the subgrid eddies and the instantaneous fields of the diagnosed eddy forcing \cite{Bolton2019}. These methods can be accurate but contain many tunable parameters, obstructing their physical interpretation. The approach proposed by \citeA{zanna2020data} (hereafter ZB20) provided an alternative to both the traditional bulk approach and black-box machine-learning modeling by enabling the discovery of a closed-form equation for the eddy parameterization directly from data. 

ZB20 parameterized the mesoscale eddy momentum fluxes through a simple interpretable expression that has strong ties with physics-based gradient models \cite{anstey2017deformation, zanna2020data, jakhar2023learning, khani2023gradient}. The ZB20 parameterization was trained on the data of the primitive equation ocean model MITgcm \cite{marshall1997finite} and accurately predicts the eddy fluxes, including the upgradient fluxes (backscatter),  with a skill comparable to neural-network approaches. Once implemented online, the ZB20 parameterization improves the representation of the mean flow and energy distributions \cite{zanna2020data}. However, their online simulations were limited to a simple one-layer shallow water model.  

In this work, we implement the ZB20 subgrid parameterization into the primitive-equation GFDL MOM6 ocean model \cite{adcroft2019gfdl}. In Section \ref{sec:ocean_model}, we describe the ocean model in the adiabatic limit, governed by the stacked shallow water equations \cite{adcroft2006methods}. In Section \ref{sec:subgrid_parameterizations}, we describe the ZB20 parameterization and propose modifications via filtering schemes that reduce the resulting grid-scale numerical instabilities and improve the large-scale KE backscatter. 
In Section \ref{sec:comparison_models}, we test the ZB20-based parameterizations in an idealized ocean configuration, Double Gyre, and show a reduction of biases in the mean state and energetics. In Section \ref{sec:sensitivity_studies}, we evaluate the ZB20-based parameterizations for a range of resolutions and show that they do not require retuning of the free parameter. This is mainly a consequence of the scale-awareness (or resolution-awareness) of the free parameter, which is solely a function of the grid spacing. We finally describe tests of the parameterization in a more complex configuration, NeverWorld2 \cite<NW2,>{marques2022neverworld2}, with a cross-equatorial basin and Southern Ocean re-entering channel in Section \ref{sec:NW2}. 
We find a more pronounced bias reduction in energy distributions and mean state, compared to the simpler Double Gyre configuration.
However, our experimentation in the NW2 configuration revealed the need for additional modifications to improve the numerical stability properties of the parameterization, similarly to other backscatter schemes \cite{yankovsky2023}. 

\section{Ocean model} \label{sec:ocean_model}

We use the GFDL MOM6 ocean model \cite{adcroft2019gfdl} in an adiabatic limit with no buoyancy forcing. This allows us to test the direct impact of the new parameterization in idealized settings of a primitive equation model.

The equations of motion are given by the stacked shallow water equations with constant density in each layer \cite{marques2022neverworld2, zhang2023implementation}:
\begin{gather}
    \partial_t \mathbf{u}_k + \frac{f + \zeta_k}{h_k} \hat{\mathbf{z}} \times  (h_k \mathbf{u}_k) + \nabla K_k + \nabla M_k = \mathbf{F}_k + \mathbf{V}_k + \mathbf{S}_k, \label{eq:gov_momentum} \\
    \partial_t h_k + \nabla \cdot (\mathbf{u}_k h_k) = 0, \label{eq:gov_mass}
\end{gather}
where $k$ is the index of the vertical fluid layer, equal to $1$ for the surface layer and to $\mathcal{K}$ for the bottom layer; $\mathbf{u}_k=(u_k,v_k)$ is the horizontal velocity, where $u_k$ and $v_k$ are zonal and meridional velocities; $h_k$ is the layer thickness; $f$ is the Coriolis parameter; $\zeta_k=\partial_x v_k-\partial_y u_k$ is the vertical component of the relative vorticity; $\nabla=(\partial_x, \partial_y)$ is the horizontal gradient operator and $\nabla \cdot$ is the horizontal divergence operator, where $\partial_x$ and $\partial_y$ are partial derivatives along zonal and meridional directions; $\hat{\mathbf{z}}$ is the unit vector pointing upward; $\hat{\mathbf{z}} \times \mathbf{u}_k = (-v_k, u_k)$ is the cross product; $K_k=(1/2) \mathbf{u}_k\cdot \mathbf{u}_k$ is the KE per unit mass. The Montgomery potential is given by $M_k=\sum_{l=1}^k g'_{l-1/2} \eta_{l-1/2}$, where $\eta_{k+1/2}=-H+\sum_{n=k+1}^{\mathcal{K}} h_n$ is the interface position between layers $k$ and $k+1$ and $H(x,y) \geq 0$ is the ocean depth; 
$g'_{k+1/2}=g(\rho_{k+1}-\rho_{k})/\rho_0$ is the reduced gravity, where $\rho_k$ is the density of the fluid layer, $\rho_0$ is the reference density and $g$ is the gravitation acceleration. The equations of motion in the horizontal orthogonal curvilinear coordinates are discussed in \citeA{adcroft2019gfdl}. $\mathbf{F}_k$ represents the wind stress and bottom drag and  $\mathbf{S}_k$ is a subgrid momentum parameterization. $\mathbf{V}_k$ is a biharmonic Smagorinsky model
(subsequently, we omit index $k$ for brevity), with 
a  viscosity $\nu_4 = C_S \Delta^4 \sqrt{\widetilde{D}^2+D^2}$, where $C_S$ is the non-dimensional Smagorinsky coefficient,  $\Delta$ is the grid spacing, $\widetilde{D}=\partial_x u - \partial_y v$ is the stretching deformation, and $D=\partial_y u + \partial_x v$ is the shearing deformation. 
We refer the reader to  \citeA{griffies2000biharmonic} for more details regarding the form and implementation of the biharmonic operator used in MOM6.

\section{Subgrid parameterizations} 
Ocean models at a coarse grid resolution have strong biases in the representation of the mean flow and energetics \cite{hallberg2013using, hewitt2020resolving}. They can be corrected to some degree by parameterizing the effect of the unresolved (subgrid) mesoscale eddies.
In this section, we describe how to diagnose the effect of subgrid mesoscale eddies on the resolved flow in the momentum equation using a spatial filtering approach \cite<Large Eddy Simulation, LES,>{sagaut2006large, fox2008can, bachman2017scale}. We then describe our implementation of the ZB20 parameterization of the subgrid mesoscale eddies and different baselines. The buoyancy fluxes and their parameterizations \cite{gent1990isopycnal} are absent in all model calculations documented in the paper.

\label{sec:subgrid_parameterizations}
\subsection{Subgrid momentum forcing}
The subgrid mesoscale eddies produce the following subgrid momentum forcing acting on the resolved eddies \cite{zanna2020data}
\begin{equation}
    \mathbf{S} = (\overline{\mathbf{u}} \cdot \overline{\nabla} ) \overline{\mathbf{u}} - \overline{(\mathbf{u} \cdot \nabla) \mathbf{u}},
    \label{eq:subgrid_forcing}
\end{equation}
where $\mathbf{u}$ is the velocity field of the high-resolution model, $\overline{(\cdot)}$ is a spatial filtering and coarse-graining operator. Here  $(\overline{\mathbf{u}} \cdot \overline{\nabla} ) \overline{\mathbf{u}}$ is the numerical approximation of the advection operator on a coarse grid with the scheme used in MOM6. Specifically, we employ the  \citeA{sadourny1975dynamics} energy-conserving scheme formulated in vector-invariant form using the identity $(\mathbf{u} \cdot \nabla) \mathbf{u} = (\zeta / h) \hat{\mathbf{z}} \times  (h\mathbf{u}) + \nabla K$. The subgrid forcing (Eq. \eqref{eq:subgrid_forcing}) modifies the governing equations of the coarse ocean model as shown in Eq. \eqref{eq:gov_momentum}. Note that in the LES approach, we should use $\overline{\mathbf{u}}$ whenever referring to the solution of the coarse model, but here, for brevity, we omit this notation everywhere apart from in Eq. \eqref{eq:subgrid_forcing}. To enable computations with a coarse ocean model, we should represent the subgrid forcing as a function of the resolved flow, referred to as a \textit{parameterization}.


\subsection{Zanna-Bolton parameterization (ZB20)} \label{sec:ZB}
In this section, we describe the original ZB20 parameterization and two filtered modifications, referred to as filtered ZB20 parameterizations. All three variants are referred to as ZB20-based parameterizations.

The original ZB20 parameterization for subgrid momentum forcing is given by
\begin{equation}
     \mathbf{S} = \begin{pmatrix}
    S_x \\
    S_y
    \end{pmatrix} = \nabla \cdot \mathbf{T} = \nabla \cdot 
    \begin{pmatrix}
        T_{xx} & T_{xy} \\
        T_{xy} & T_{yy} 
    \end{pmatrix} = 
    \begin{pmatrix}
        \partial_x T_{xx} + \partial_y T_{xy} \\
        \partial_x T_{xy} + \partial_y T_{yy} 
    \end{pmatrix} \label{eq:ZB_momentum_forcing}. 
\end{equation}
The parameterization was discovered, using a machine learning algorithm, from data generated from baroclinic ocean simulations \citeA{zanna2020data}. 
The stress tensor $ \mathbf{T}$ is given by
\begin{equation}
    \mathbf{T}(\zeta, D, \widetilde{D}) = \kappa_{BC}
    \underbrace{
    \begin{bmatrix}
        - \zeta D & \zeta \widetilde{D} \\
        \zeta \widetilde{D} & \zeta D 
    \end{bmatrix}}_{\mathbf{T}_{\mathrm{d}}\text{, deviatoric stress}}
    + 
    \frac{\kappa_{BC}}{2}
    \underbrace{
    \left( \zeta^2 +D^2 + \widetilde{D}^2 \right)
    \begin{bmatrix}
        1 & 0 \\
        0 & 1
    \end{bmatrix}}_{\mathbf{T}_{\mathrm{I}}\text{, isotropic stress}}. \label{eq:ZB_momentum_flux}
\end{equation}
The parameterization is applied independently in every model layer.
Here, we follow the approach of \citeA{anstey2017deformation} and gradient models studies \cite{chen2003physical, meneveau2000scale} to relate the free coefficient $\kappa_{BC}$ to the area of a coarse grid box:
\begin{equation}
    \kappa_{BC} = - \gamma \Delta_x \Delta_y \leq 0, \label{eq:kappa_bc}
\end{equation}
where $\Delta_x$ and $\Delta_y$ are local grid spacings along $x$ and $y$ directions, respectively, and $\gamma \approx 1$ is a tunable non-dimensional parameter. In 2D incompressible fluids, the ZB20 parameterization (Eq. \eqref{eq:ZB_momentum_flux}) is equivalent to the nonlinear gradient model, which is given by $\mathbf{T} = - l^2 (\nabla \mathbf{u}) \cdot (\nabla \mathbf{u})^{\dag}$, where $\nabla \mathbf{u}$ is the velocity gradient tensor, $l$ is the filter width, $\dag$ is the matrix transpose and $\cdot$ is the matrix multiplication; see Eq. (59) in \citeA{anstey2017deformation} for details. However, as opposed to the nonlinear gradient model, the ZB20 parameterization excludes the explicit dependence on the horizontal divergence ($\sigma = \partial_x u + \partial_y v$) in stacked shallow water equations (see, for example, \citeA{zanna2020data} for further discussion on the difference between ZB20 and the nonlinear gradient model). Based on limited numerical experiments in the NW2 configuration, the additional terms in the nonlinear gradient models, which explicitly depend on $\sigma$, have a destabilizing effect on the simulations (not shown).


We refer the reader to \ref{appendix:model_formulation} for details of the numerical discretization of Eqs. (\ref{eq:ZB_momentum_forcing}-\ref{eq:ZB_momentum_flux}), accounting for the curvilinear coordinates and varying layer thickness in MOM6. 
 For interested readers, we note that, as opposed to the original ZB20 model, the implemented ZB20 parameterization may spuriously predict non-zero accelerations for a state of solid body rotation ($\zeta =\mathrm{const}$, $D=\widetilde{D}=0$).  
These non-zero accelerations are caused by the spatially varying parameterization coefficient $\kappa_{BC}$, leading to a non-zero divergence of the isotropic stress tensor ($\nabla \cdot (\kappa_{BC} \mathbf{T}_{\mathrm{I}}) \neq 0$).
 We anticipate that the non-zero accelerations will be small when the coefficient $\kappa_{BC}$ varies over large spatial scales, as it is the case for the ZB20-based parameterizations in the current implementation. 
Other spatially varying coefficients, such as the layer thickness in the stress divergence expression (Eq. \eqref{eq:ZB_thickness}) or the Coriolis parameter in the attenuation function (Eq. \eqref{eq:Klower_attenuation}, to be introduced later) can lead to a non-zero divergence of the isotropic stress tensor as well. 

\subsubsection{Low-pass filtering of the stress tensor (ZB20-Smooth)} \label{sec:ZB_smooth}
Incorporation of the ZB20 parameterization, which is meant to dissipate energy at small scales,  can generate numerical noise near the grid scale. 
It happens because the dissipation near the grid scale is not a hard constraint of the parameterization.  Specifically, the energetic contribution of the deviatoric stress is zero after integration by parts  ($\mathbf{T}_{\mathrm{d}}:(\nabla \mathbf{u}) =0$, see Eq. \eqref{eq:ZB_energy_flux} in \ref{appendix:model_formulation}) while the energetic contribution of the isotropic stress ($\mathbf{T}_{\mathrm{I}}:(\nabla \mathbf{u})$) is not sign-definite \cite{zanna2020data}. 
Dissipation at the grid scale is often enforced in gradient-based parameterizations by projecting the predicted stress tensor onto the dissipative direction \cite{bouchet2003parameterization, balarac2013dynamic, vollant2016dynamic}. However, this approach is unsuitable because it would remove an important effect of the KE backscatter on large scales. Thus, we suggest removing the contribution of the ZB20 parameterization on the grid scale eddies by low-pass filtering the stress tensor. The filtered stress tensor will represent the KE backscatter. Filtering is widely used in mesoscale eddy parameterizations to suppress the numerical noise and increase the spatial scale of the KE backscatter \cite{grooms2015numerical, juricke2019ocean, bachman2019gm, perezhogin2019stochastic}.

We consider a low-pass convolutional filter defined on $3 \times 3$ spatial stencil (``trapezoidal filter'' in \citeA{san2014dynamic}), which is applied in every fluid layer independently:
\begin{equation}
    \mathcal{G} = \frac{1}{16}
    \begin{pmatrix}
        1 & 2 & 1 \\
        2 & 4 & 2 \\
        1 & 2 & 1
    \end{pmatrix}. \label{eq:G_filter}
\end{equation}
The presented filter has the smallest spatial stencil among filters, completely removing the grid harmonics. It will be used as a building block for proposing our filtered ZB20 parameterizations.

The ZB20 parameterization with low-pass filtered stress tensor (hereafter, ZB20-Smooth) is given by:
\begin{equation}
    \mathbf{S} = \nabla \cdot  G(\mathbf{T}),
\end{equation}
where $G=\mathcal{G}^N$ is the low-pass filter which is applied to every component of the stress tensor $\mathbf{T}$ (Eq. \eqref{eq:ZB_momentum_flux}), and $N$ is the number of filtering passes of $\mathcal{G}$ (Eq. \eqref{eq:G_filter}). We choose $N=4$ similarly to \citeA{juricke2019ocean} (see \ref{sec:additional_sensitivity} for sensitivity to the choice of $N$). We implement the filtering using the marching-halo algorithm; that is, 4 filter iterations are performed within a single MPI exchange. The computational cost of the original ZB20 parameterization is $2.5 \%$ of the ocean model runtime in the NW2 configuration, while the filtered parameterization requires $4\%$ of the runtime. Note that filters more appropriate for tensor elements were presented in \citeA{aluie2019convolutions}. 

\subsubsection{High-pass filtering of the velocity gradients (ZB20-Reynolds)} \label{sec:ZB_reynolds}
We consider an additional filtering scheme, which can also enhance the KE backscatter.

\citeA{perezhogin2023subgrid} show that the Reynolds stress is responsible for the KE backscatter. The Reynolds stress represents the effect of the eddy-eddy interactions on the mean flow, through the \citeA{germano1986proposal} decomposition of the subgrid stress. To isolate the effect of eddy-eddy interaction from the Reynolds stresses, we propose a modification of the ZB20 parameterization by using a high-pass filter on the velocity gradients. 
Therefore, the modified parameterization (hereafter, ZB20-Reynolds) can be expressed as 
\begin{equation}
    \mathbf{S} = \nabla \cdot  G(\mathbf{T}(\zeta', D', \widetilde{D}')), \label{eq:ZB-Reynolds}
\end{equation}
where $G=\mathcal{G}^N$ is the low-pass filter and $(\cdot)'=I-G$ is the high-pass filter with $I$ being the identity operator. The computation of the stress tensor in Eq. \eqref{eq:ZB-Reynolds} is done as follows. First, the velocity gradients $\zeta$, $D$, $\widetilde{D}$ are high-pass filtered to obtain $\zeta'$, $D'$, $\widetilde{D}'$, respectively. Then we compute the stress tensor $\mathbf{T}(\zeta', D', \widetilde{D}')$ according to Eq. \eqref{eq:ZB_momentum_flux} using high-pass filtered fields as inputs. Finally, the stress tensor $\mathbf{T}$ is low-pass filtered with the filter $G$ to separate the parameterization tendency from the grid scale. For consistency with the previous section, we choose $N=4$. This additional filtering results in an increase in the computational cost of the parameterization, which is $6\%$ of the ocean model runtime in the NW2 configuration.

\subsection{Baseline momentum parameterizations}
We consider multiple backscatter parameterizations as baselines. 

The first baseline is the KE backscatter of \citeA{jansen2015energy} (referred to as JHAH15), already tested in MOM6 \cite{jansen2019toward}. The JHAH15  parameterization mainly represents the reinjection of KE energy originally dissipated with the biharmonic Smagorinsky model. The backscatter of the subgrid KE is parameterized using a negative Laplacian viscosity model (anti-viscosity). The negative viscosity coefficient is informed by a local equation for vertically-averaged subgrid KE.
We will also consider an updated version of the JHAH15 parameterization, from \citeA{yankovsky2023} for multi-layer models in Section~\ref{sec:NW2}.

The second baseline parameterization is a deep-learning convolution neural network (CNN) model of \citeA{guillaumin2021stochastic} (referred to as GZ21). It predicts the subgrid forcing using horizontal velocities and was trained on data from a coupled climate simulation. It was implemented in MOM6 \cite{zhang2023implementation}, together with a biharmonic Smagorinsky model. The GZ21 parameterization energizes the resolved eddies efficiently and consequently can be tuned to increase the KE of the coarse model up to the KE of the high-resolution model. 

\section{Experiments in Double Gyre configuration} \label{sec:comparison_models}

\begin{figure}[h!]
\centerline{\includegraphics[width=1.0\textwidth]{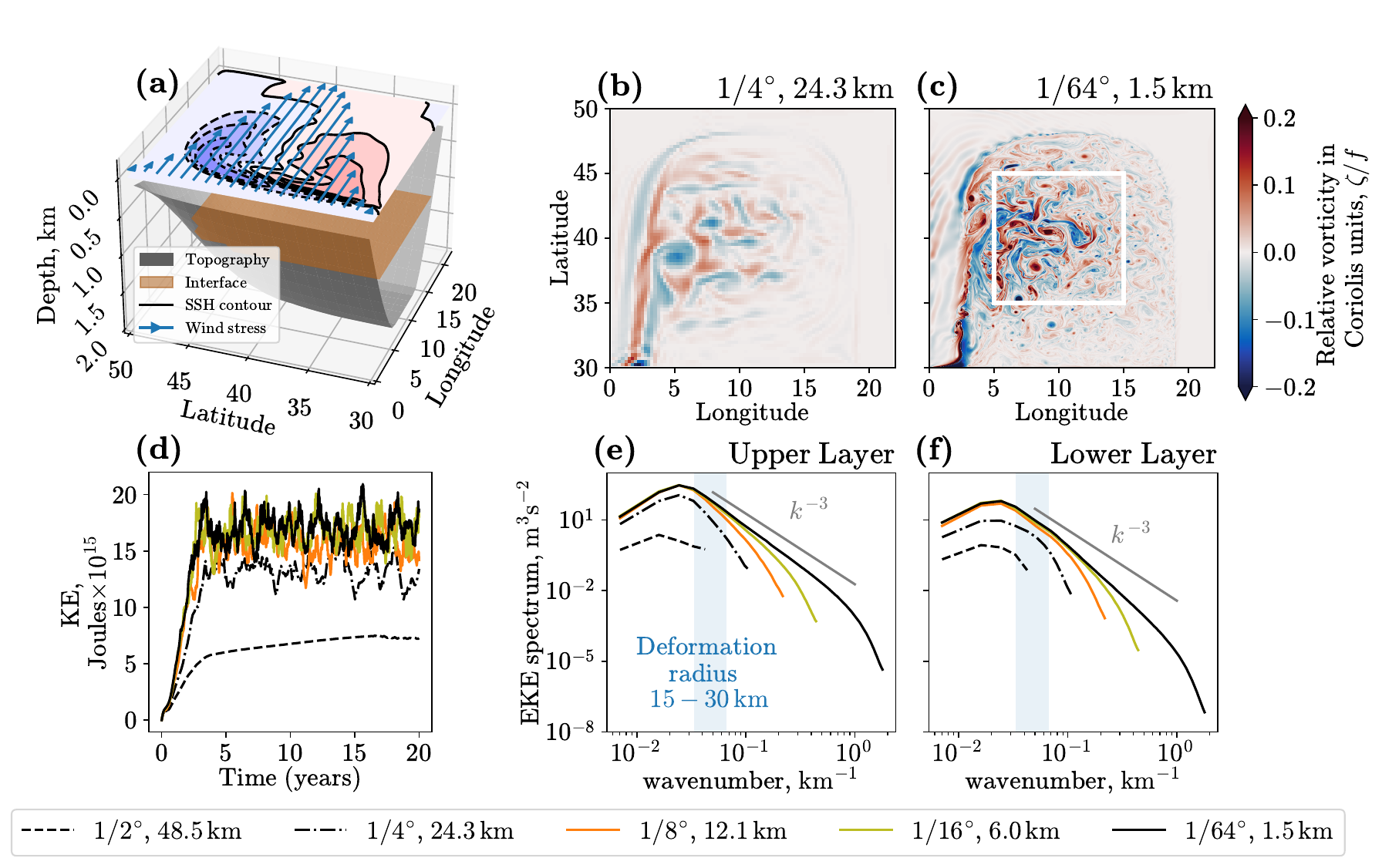}}
\caption{(a) Schematic of the MOM6 Double Gyre configuration; (b) snapshot of relative vorticity in the upper layer in the 1/4$^\circ$ simulation and (c) 1/64$^\circ$ simulation; (d) timeseries of total kinetic energy (summed over layers) for the different resolutions. The spatial spectra of the eddy kinetic energy (EKE) as a function of wavenumbers in the (e) upper and (f) lower fluid layers. The spectrum is computed within the area enclosed by the white rectangle. The wavenumbers corresponding to the deformation radius ($r_d \approx f^{-1}\sqrt{g' h_1 h_2 (h_1+h_2)^{-1}}$), given by $1/r_d$, are shaded in blue. }
\label{fig:Double-Gyre}
\end{figure} 


We first test the subgrid parameterizations in the MOM6 Double Gyre configuration described in Section 3.1 of \citeA{zhang2023implementation}. 
We use their setting but double the integration time (20 years) with a single ensemble member.
The model has two fluid layers, initially at rest. Momentum input is via the wind and dissipation by the bottom friction.
The computational domain is on a spherical grid with a bowl topography (Figure \ref{fig:Double-Gyre}(a)). The Smagorinsky coefficient in all experiments is set to $C_S=0.06$, similar to \citeA{jansen2015energy, jansen2019toward, zhang2023implementation}; see \ref{sec:additional_sensitivity} for sensitivity to $C_S$. 

The unparameterized simulations ($\mathbf{S}=0$) for a range of resolutions (1/64$^\circ$, 1/16$^\circ$, 1/8$^\circ$, 1/4$^\circ$, 1/2$^\circ$) are shown in Figure \ref{fig:Double-Gyre}.
The high-resolution model ($1/64^\circ$) has a grid spacing ($\approx 1.5 \, \mathrm{km}$) that is 10 times smaller than the Rossby deformation radius ($15 \, \mathrm{km}-30 \, \mathrm{km}$), and consequently, it directly simulates the mesoscale eddies \cite{hallberg2013using}. Coarse ocean models with a grid spacing in a range from  $1/2^\circ$ ($\approx50 \, \mathrm{km}$) to $1/8^\circ$ ($\approx12 \, \mathrm{km}$) barely resolve the Rossby deformation radius. These models have a reduced eddy kinetic energy (EKE) spectrum compared to the high-resolution model ($1/64^\circ$), with the coarsest models ($1/2^\circ$ and $1/4^\circ$) failing to capture the spectrum at all spatial scales, see Figure \ref{fig:Double-Gyre}(e,f).



\begin{table}[h!]
\begin{center}
\begin{tabular}{c|ccc}
 & ZB20 & ZB20-Smooth & ZB20-Reynolds \\
\hline
$\gamma$ & $0.5$ & $1.0$ & $2.0$
\end{tabular}
\end{center}
\caption{Values of the default scaling coefficient $\gamma$ in the different ZB20-based parameterizations.}
\label{tab:default_coefficient}
\end{table}

In this section, we analyze the impact of subgrid parameterizations, in particular, on improving biases in energetics and the mean states in the coarse resolution model with horizontal grid spacing of $1/4^\circ$.
The values of the scaling coefficient $\gamma$ of the ZB20-based parameterizations used in this section are reported in Table \ref{tab:default_coefficient}. The coefficient $\gamma$ is larger for the filtered than for the unfiltered versions of the ZB20 parameterization, since filtering reduces the magnitude of the subgrid forcing. The subgrid parameterizations JHAH15, GZ21 and ZB20-Reynolds are tuned to approximately match the total KE, and the ZB20-Smooth parameterization is tuned to match the APE. Sensitivity to the scaling coefficient and performance at multiple resolutions are discussed in Section \ref{sec:sensitivity_studies}.  

\subsection{Eddy kinetic energy (EKE) spectrum}
\begin{figure}[h!]
\centerline{\includegraphics[width=1.0\textwidth]{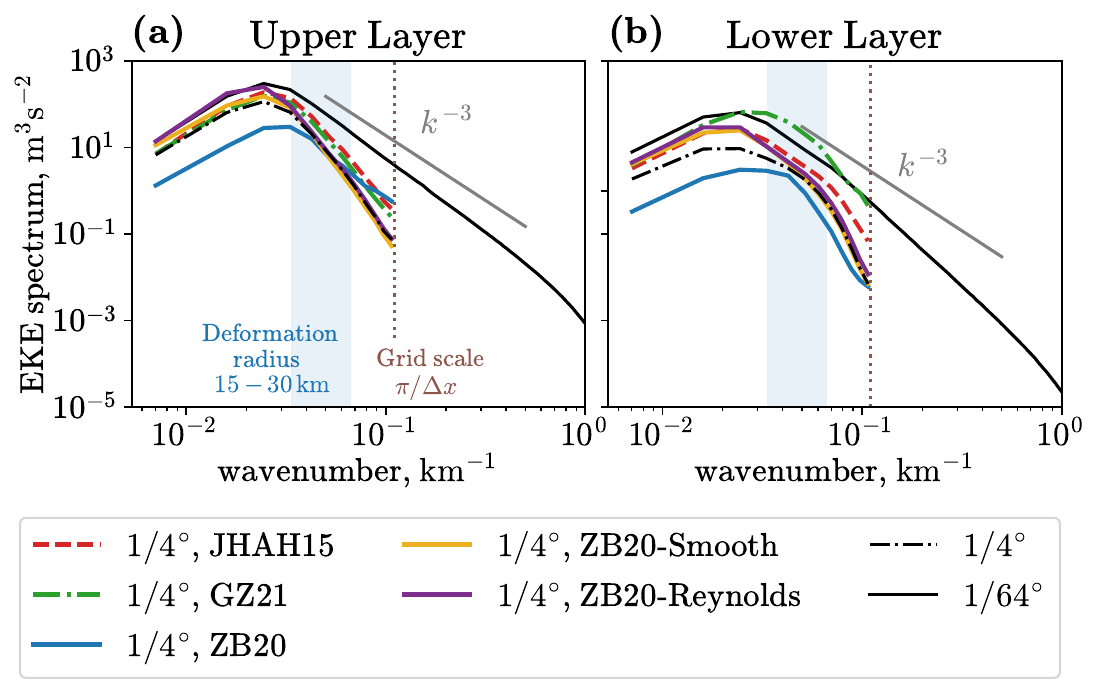}}
\caption{The eddy kinetic energy (EKE) spatial spectrum as a function of wavenumbers for the last 10 years of the simulation: (a) upper layer, (b) lower layer. The unparameterized model ($1/4^\circ$, dotted dashed line) is compared to parameterized models (JHAH15 in red, GZ21 in green, ZB20 in blue, ZB20-Smooth in yellow, ZB20-Reynolds in violet) and high-resolution simulation ($1/64^\circ$, black line). The spectrum is computed in the region indicated by the white rectangle in Figure \ref{fig:Double-Gyre}.}
\label{fig:Spectrum}
\end{figure} 
The eddy kinetic energy (EKE) spectrum is one of the metrics that coarse unparameterized ocean models fail to reproduce (Figure \ref{fig:Double-Gyre}(e,f)).
The EKE spectrum is defined as a time-averaged spatial power spectrum of the eddy velocities $\mathbf{u}' = \mathbf{u} - \overline{\mathbf{u}}^t$, where $\overline {(\cdot)}^{t}$ is a time-average over the last 10 years of the simulations. All spectra in the Double Gyre configuration are computed in the white rectangle shown in Figure \ref{fig:Double-Gyre}(c) using a 2D Fourier transform with a Hann window and linear detrending \cite<using xrft package,>{takaya_uchida_2023_7621857}. Throughout the paper, we ignore the contribution of the thickness $h$ in the definition of spectral properties related to kinetic energy.

The EKE spectra for runs with different subgrid parameterizations are shown in Figure \ref{fig:Spectrum}. The ZB20 parameterization without filters has a build-up of energy near the grid scale, i.e. numerical noise (Figure \ref{fig:Spectrum}(a)), which results in the deterioration of the EKE spectrum at large scales in both fluid layers. The proposed filtering techniques (ZB20-Smooth and ZB20-Reynolds) allow us to attenuate the grid scale noise generation. 
The ZB20-Smooth and ZB20-Reynolds parameterizations improve the EKE spectrum at large scales by increasing it compared to the unparameterized model, reaching levels closer to the high-resolution model. The ZB20-Reynolds parameterization is more efficient in energizing eddies in the upper layer. 
The JHAH15 and GZ21 baseline parameterizations are more efficient in energizing eddies near the deformation scale. None of the tested subgrid parameterizations reproduce the EKE spectrum across all spatial scales.

\subsection{Subgrid kinetic energy transfer} \label{sec:transfer}
\begin{figure}[h!]
\centerline{\includegraphics[width=1.0\textwidth]{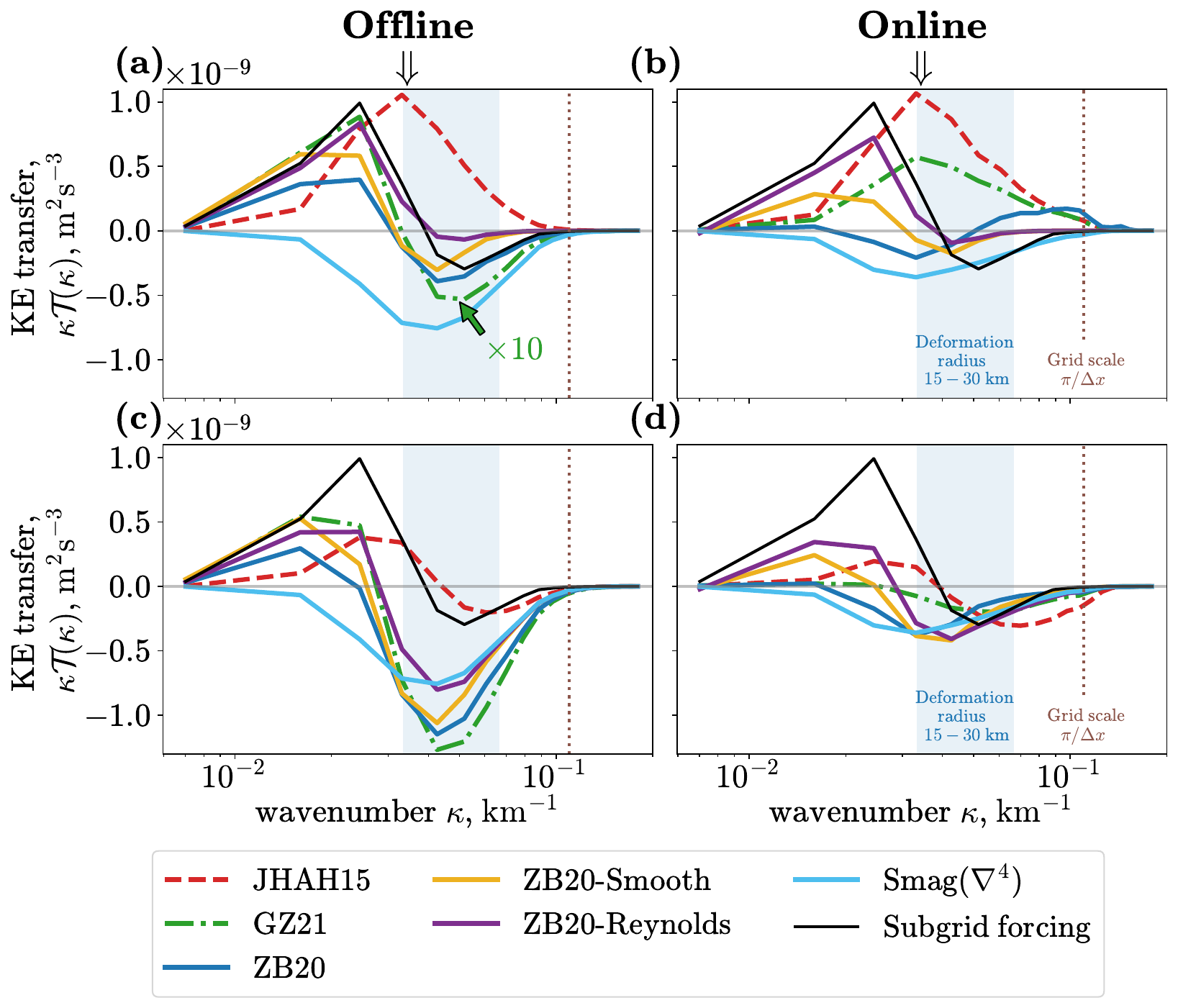}}
\caption{
Kinetic energy transfer in offline (left column) and online (right column) analysis.  
The target resolution for offline analysis is $1/4^\circ$ and the online performance of the parameterizations is computed for simulations at $1/4^\circ$ resolution. 
(a) Offline contribution from each subgrid parameterization (GZ21 in green, ZB20 in blue, ZB20-Smooth in yellow, ZB20-Reynolds in violet, negative viscosity part of JHAH15 in red) and a biharmonic Smagorinsky model ``Smag($\nabla^4$)'' in light blue.  
The black line is the subgrid forcing, Eq. \eqref{eq:subgrid_forcing}. 
(b) The contribution of each parameterization online. (c) and (d): same as (a) and (b) but showing the combined contribution of each parameterization with a biharmonic Smagorinsky model offline and online, respectively. The offline analysis is performed on fields of the high-resolution model ($1/64^\circ$) filtered and coarse-grained to $1/4^\circ$ resolution. The spectra are computed over the last 10 years of the simulation for the upper fluid layer, in the region enclosed by the white rectangle shown in Figure \ref{fig:Double-Gyre}. The KE transfer of GZ21 in panels (a) and (c) is multiplied by 10 for convenience.}

\label{fig:SGS_Spectrum}
\end{figure}

In this section, we analyze the subgrid energy transfer in coarse parameterized models to explain the shape of the EKE spectrum. The energy transfer is computed both offline and online because the accuracy of the parameterization can change once it is included in the coarse-grid model \cite{meneveau2000scale, ross2022benchmarking}. Note that the offline analysis evaluates the subgrid parameterization on the filtered and coarse-grained snapshots of the high-resolution model, while the online analysis evaluates the parameterization once the parameterized ocean model is integrated over time.

The KE transfer spectrum of the subgrid forcing or parameterization $\mathbf{S}=(S_x,S_y)$ is given by
\begin{equation}
    \mathcal{T}(\kappa_x, \kappa_y) = Re\left[\mathcal{F}(u)^* \mathcal{F}(S_x) + \mathcal{F}(v)^* \mathcal{F}(S_y)\right],
\end{equation}
where $\kappa_x$ and $\kappa_y$ are zonal and meridional wavenumbers, respectively, $\mathcal{F}$ is the 2D Fourier transform, $^*$ is the complex conjugate, and $Re$ is the real part. We integrate the two-dimensional transfer spectrum over circles ($\kappa_x^2+\kappa_y^2=\kappa^2$) to obtain the isotropic transfer spectrum $\mathcal{T}(\kappa)$. We compute the subgrid forcing according to Eq. \eqref{eq:subgrid_forcing}, where the filtering operator $\overline{(\cdot)}$ is defined as the Gaussian filter implemented in the GCM-filters package \cite{grooms2021diffusion, loose2022gcm} followed by coarse-graining. The filtering and coarse-graining operations are applied to the $1/64^\circ$ model output towards a $1/4^\circ$ resolution. The width of the Gaussian filter is chosen to be proportional to the grid spacing of the coarse resolution model by introducing a filter-to-grid width ratio \cite<FGR,>{ghosal1996analysis, lund1997use, chow2003further, perezhogin2023subgrid}. The FGR is not a coefficient used for the online implementation of the parameterization; it is only a parameter used to diagnose the offline subgrid transfer. 
The KE transfer of the subgrid models in the offline analysis depends on the FGR, because, as expected, it depends on grid-scale features. 
However, the shape of the offline transfer spectrum as a function of wavenumber remains roughly unchanged, while the amplitude varies for different FGR values.
For $\mathrm{FGR}=2.5$, the offline and online KE transfer roughly coincide for the filtered ZB20 and JHAH15 parameterizations (Figure \ref{fig:SGS_Spectrum}). We chose this FGR value in our analysis,  which likely corresponds to the filter width that is effectively reproduced in online simulations by the parameterized ocean models.

The subgrid energy transfer contains an interval with small-scale energy and enstrophy dissipation ($\mathcal{T}(\kappa) < 0$) and an interval of large-scale energy backscatter ($\mathcal{T}(\kappa) > 0$), see black line in Figure \ref{fig:SGS_Spectrum}.  The unfiltered ZB20 parameterization simulates the positive and negative energy transfer relatively well in the offline analysis; see Figure \ref{fig:SGS_Spectrum}(a), blue line. However, in the online simulations, the energy transfer is reversed: the ZB20 parameterization dissipates energy at large scales and generates energy near the grid scale (Figure \ref{fig:SGS_Spectrum}(b)). This explains the emergence of the grid-scale numerical noise and deterioration of the large scales in the EKE spectrum for the simulation with the unfiltered ZB20 parameterization (Figure \ref{fig:Spectrum}). 
The filtering techniques employed for the ZB20-Smooth and ZB20-Reynolds parameterizations remove the generation of the numerical noise near the grid scale and preserve the original properties of the ZB20 parameterization. Consequently, the backscattering for these filtered parameterizations is located over the large scales (Figure \ref{fig:SGS_Spectrum}(b)). The ZB20-Reynolds parameterization appears more efficient than ZB20-Smooth in reproducing the KE backscatter.

The dissipation predicted by the filtered ZB20 parameterizations alone is insufficient to parameterize the direct cascade of enstrophy from resolved to subgrid scales because the filtering schemes diminish the effect of these parameterizations at short wavelengths. Thus, all the online experiments are performed with a biharmonic Smagorinsky model, which is effective at short wavelengths. We note that the eddy viscosity model of \citeA{leith1996stochastic} may be more suitable for parameterizing the enstrophy dissipation \cite{bachman2017scale}. The lower row in  Figure \ref{fig:SGS_Spectrum} shows a net KE transfer between the resolved and subgrid scales, which is given by a combined contribution of ZB20-based parameterizations with a biharmonic Smagorinsky model. 
The net KE transfer for the filtered ZB20 parameterizations exhibits less accurate correspondence with the diagnosed subgrid KE transfer, displaying an overestimation of dissipation and an underestimation of backscatter, both offline and online (see Figure \ref{fig:SGS_Spectrum}(c,d)). However, the efficacy of the filtered ZB20 parameterizations can be attributed to the scale separation between backscatter and dissipation effects in the net KE transfer, which is crucial for effectively energizing the resolved flow \cite{jansen2014parameterizing, bagaeva2023}.

The JHAH15 parameterization returns energy in shorter wavelengths than the filtered ZB20 parameterizations (Figure \ref{fig:SGS_Spectrum}(c,d)). We explain the increased energy density near the deformation scale for the JHAH15 parameterization (Figure \ref{fig:Spectrum}) by the effect of scale-selective backscatter, which may be seen as an advantage; however, it can also lead to numerical instabilities \cite{grooms2015numerical, bachman2019gm, juricke2019ocean, grooms2023backscatter, bagaeva2023}. Note that while the net transfer for the JHAH15 parameterization does not match the diagnosed KE transfer well, the JHAH15 parameterization is expected to be more accurate when compared to the subgrid forcing diagnosed with the cut-off LES filter \cite{perezhogin2023subgrid}. 

For the convenience of plotting, we multiply the offline KE transfer of the GZ21 parameterization by 10 in Figure \ref{fig:SGS_Spectrum}(a,c); this correction is not applied in online runs presented on panels (b,d). Note that the magnitude of offline prediction depends a lot on the FGR, while the shape of the predicted KE transfer is less FGR-sensitive and thus a more crucial metric.  The GZ21 parameterization accurately predicts the shape of the KE transfer in offline analysis (Figure  \ref{fig:SGS_Spectrum}(a)). However, like the unfiltered ZB20 parameterization, GZ21 performs differently for online experiments. The online KE transfer spectrum for the GZ21 parameterization is purely positive and lacks a dissipative region. The shape of the predicted KE transfer resembles the negative Laplacian viscosity model (Figure  \ref{fig:SGS_Spectrum}(b)), thus suggesting that GZ21 returns energy in short wavelengths and should be effective in improving the energy density near the deformation scale (Figure 2). A potential drawback of the GZ21 parameterization is the lack of scale separation between dissipation and backscatter models (Figure \ref{fig:SGS_Spectrum}(d)).

\subsection{Kinetic and potential energy} \label{sec:KE_PE_section}
\begin{figure}[h!]
\centerline{\includegraphics[width=1.0\textwidth]{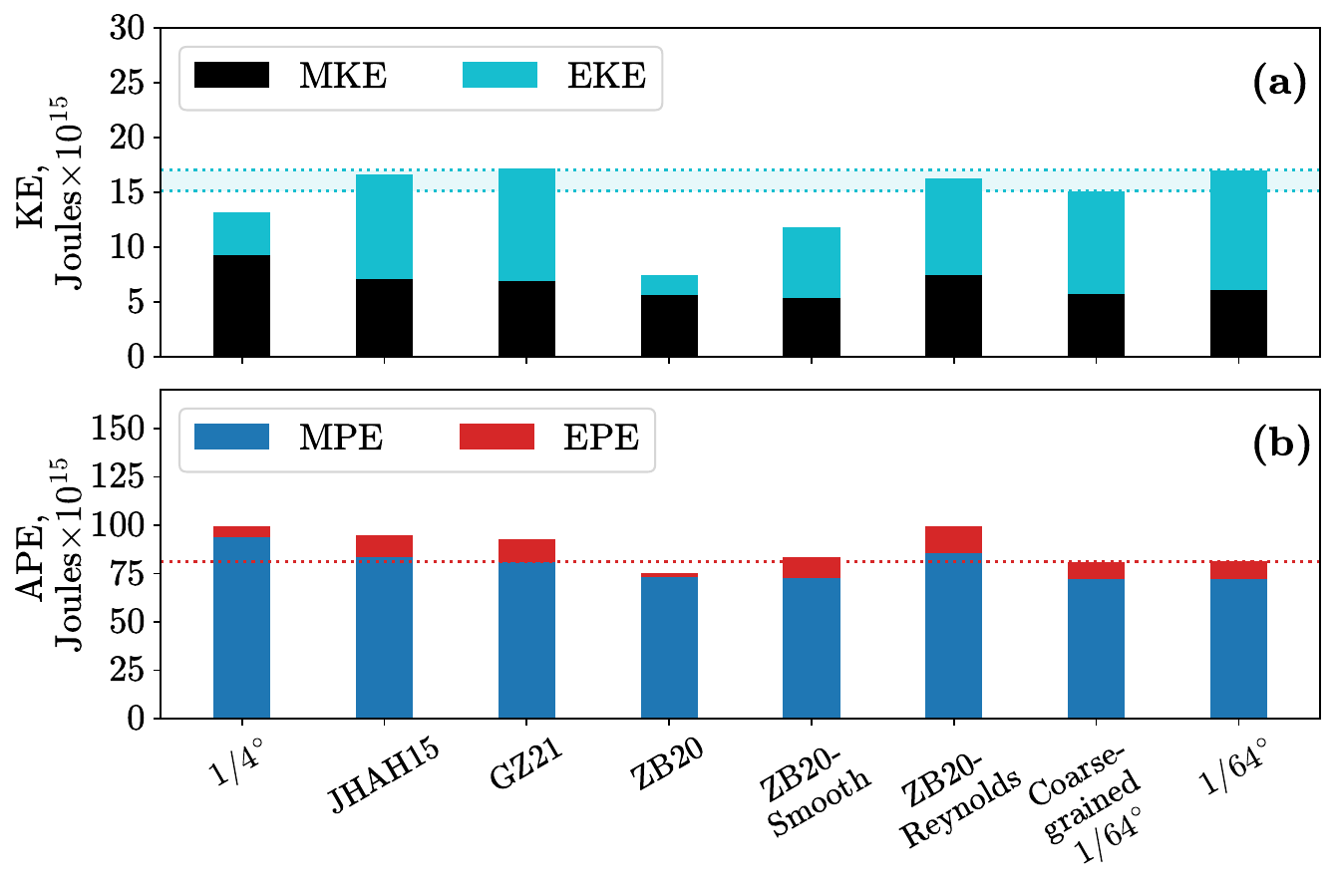}}
\caption{Energy reservoirs in numerical experiments at resolution $1/4^\circ$: (a) Kinetic energy (KE) of the mean flow (MKE) and eddies (EKE), (b) Available potential energy (APE) of the mean flow (MPE) and eddies (EPE). In both plots, energy is summed over the fluid layers or interfaces and averaged over the last 10 years. The horizontal dotted lines represent the ground truth value of the total kinetic (a) and total potential (b) energy.
}
\label{fig:KE_PE}
\end{figure} 

We compute the kinetic energy (KE) and available potential energy (APE) (\ref{appendix:KE_APE}) to address how the parameterizations affect the energy partitioning between these reservoirs. We further split each reservoir such that the KE is the sum of the mean kinetic energy (MKE), averaged over the last 10 years of the simulations, and the eddy kinetic energy (EKE). We similarly decompose the APE into the mean potential energy (MPE) and eddy potential energy (EPE).

We compare the coarse-resolution models, with and without parameterizations, to the high-resolution model and its coarse-grained output (Figure \ref{fig:KE_PE}). 
The APE of the unparametrized model ($1/4^\circ$) is too large, and its KE is too low compared to the high-resolution simulation. 
The ZB20 parameterization improves (reduces) the APE, as expected from the mesoscale parameterization 
(Figure \ref{fig:KE_PE}(b)). However, due to the generation of numerical noise, the large-scale eddies are disrupted, and the total KE is too low (Figure \ref{fig:KE_PE}(a)). Note that compared to the physical buoyancy parameterizations \cite{gent1990isopycnal}, the reduction of APE by the ZB20 parameterization appears indirectly through the change of the mean state. Directly reducing the APE would require an additional parameterization in the thickness equation \cite{loose2023diagnosing}, which we omit here. The ZB20-Smooth parameterization improves (reduces) the APE but with little change to the KE; it also reduces the MKE in agreement with the high-resolution model (Figure \ref{fig:KE_PE}(a)). The backscatter parameterizations (ZB20-Reynolds, JHAH15 and GZ21) efficiently energize the flow by increasing the KE. Additionally, they improve (reduce) the energy of the mean state (MKE and MPE). However, they are less accurate in predicting the APE than the ZB20-Smooth parameterization.

\subsection{Mean state}
\begin{figure}[h!]
\centerline{\includegraphics[width=1.0\textwidth]{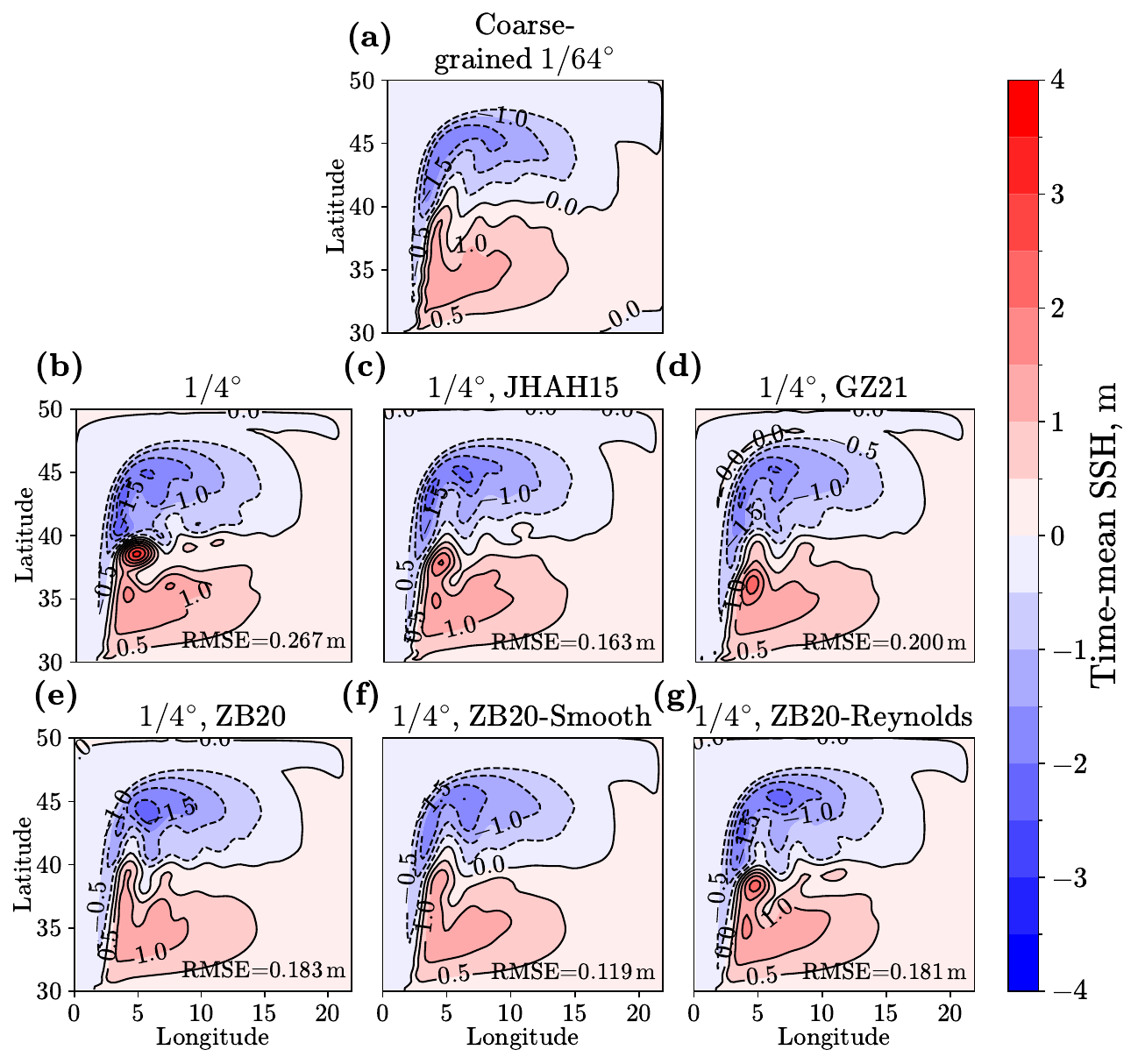}}
\caption{Sea surface height (SSH) averaged over the last 10 years for experiments at $1/4^\circ$ resolution. For every coarse model, we provide the root mean squared error (RMSE) in the time-mean SSH with respect to the coarse-grained high-resolution model shown in panel (a).}
\label{fig:ssh}
\end{figure} 

We concentrate on metrics related to the climatological mean state (Figure \ref{fig:ssh}), such as the time-mean sea surface height (SSH). The unparameterized model ($1/4^\circ$) has a strong persistent recirculation near the western boundary, absent in the high-resolution model. In all parameterized simulations (ZB20, ZB20-Smooth, ZB20-Reynolds, JHAH15, GZ21), we observe an improvement in the representation of the mean state. The persistent recirculation is less evident, and the region exhibits a meandering jet pattern similar to that simulated in the high-resolution model. The ZB20-Smooth model is the most accurate in representing the mean state, and its root mean squared error (RMSE)  is significantly lower than for other parameterized models (see Figure \ref{fig:ssh} for RMSE SSH values used to describe the mean bias in SSH). The ZB20-Smooth parameterization improves the mean state without increasing the KE (i.e., without strong backscatter). However, the most efficient backscatter parameterizations (ZB20-Reynolds, JHAH15, GZ21) have higher RMSE for SSH. By increasing the scaling coefficient $\gamma$, we can further improve the RMSE for the ZB20-Reynolds model (\ref{sec:additional_sensitivity}). However, the value is constrained by numerical stability, and here, we find the optimal coefficient $\gamma=2.8$ to be on the boundary of the stability region. 

\begin{figure}[h!]
\centerline{\includegraphics[width=1.0\textwidth]{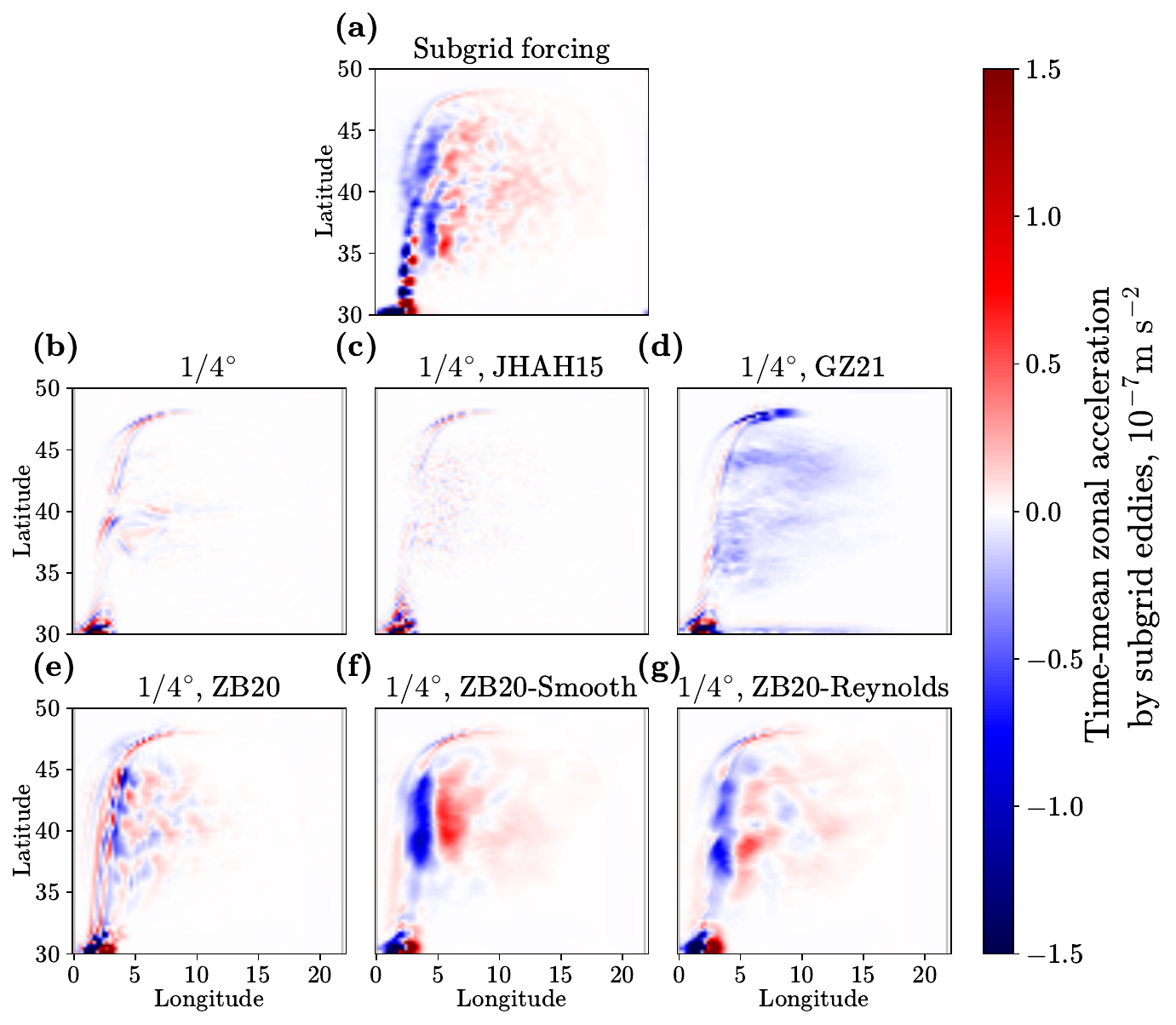}}
\caption{Analysis of the eddy-mean flow interaction following \citeA{hughes2001eddy}: 
the 10-year mean zonal acceleration in the upper fluid layer produced by the subgrid eddies ($\overline{S}^t_x$). (a) The zonal component of the subgrid forcing (Eq. \eqref{eq:subgrid_forcing}), diagnosed from the $1/64^\circ$ model by filtering and coarse-graining. Panels (b)-(g): the zonal acceleration, in online simulations, produced by combining each parameterization with a biharmonic Smagorinsky model.} 
\label{fig:eddy_mean_interaction}
\end{figure} 

\subsection{Eddy-mean flow interaction}
The two previous sections show that the parameterizations that are best in reproducing the KE backscatter do not demonstrate optimal performance in reproducing the mean SSH and APE. According to \citeA{moser2021statistical}, the time-mean subgrid stress is another property of the subgrid eddies that affects the mean flow prediction. In geophysical fluid flows, a similar effect is known as eddy-mean flow interaction and has been analyzed in several studies \cite{andrews1976planetary, hoskins1983shape, greatbatch1998exploring, wardle2000representation, hughes2001eddy, kamenkovich2009role, qiu2010eddy, greatbatch2010ocean, greatbatch2010transport, waterman2011eddy}. 
Following \citeA{hughes2001eddy}, we show the 10-year averaged zonal acceleration produced by the subgrid eddies in the upper layer (Figure \ref{fig:eddy_mean_interaction}). 
The subgrid forcing amplifies the resolved jet by accelerating the jet current extension eastward (longitude $>5^\circ$) and decelerating the jet in the separation region (longitude $<5^\circ$), see Figure \ref{fig:eddy_mean_interaction}(a). A similar pattern was shown in \citeA{zanna2020data}; see their Figures S2 and S4. The time-mean contribution of the eddy-viscosity models (biharmonic Smagorinsky and JHAH15) is too small compared to the diagnosed subgrid forcing (Figure \ref{fig:eddy_mean_interaction}(b,c)), as expected from \citeA{moser2021statistical}. The GZ21 parameterization produces westward accelerations in most of the domain and thus disagrees with the diagnosed subgrid forcing (Figure \ref{fig:eddy_mean_interaction}(d)). The ZB20-Smooth and ZB20-Reynolds parameterizations reproduce the acceleration pattern of the subgrid forcing most accurately, with the ZB20-Smooth having the largest accelerations (Figure \ref{fig:eddy_mean_interaction}(f,g)). The zonal acceleration produced by the unfiltered ZB20 parameterization is smaller and less accurate compared to the filtered models (Figure \ref{fig:eddy_mean_interaction}(e)). The success of the ZB20-based parameterizations in improving the mean state and the APE appears to be related to their effect on the time-mean zonal acceleration \cite{anstey2017deformation}.

\section{Sensitivity study and scale awareness} \label{sec:sensitivity_studies}
\begin{figure}[h!]
\centerline{\includegraphics[width=1.0\textwidth]{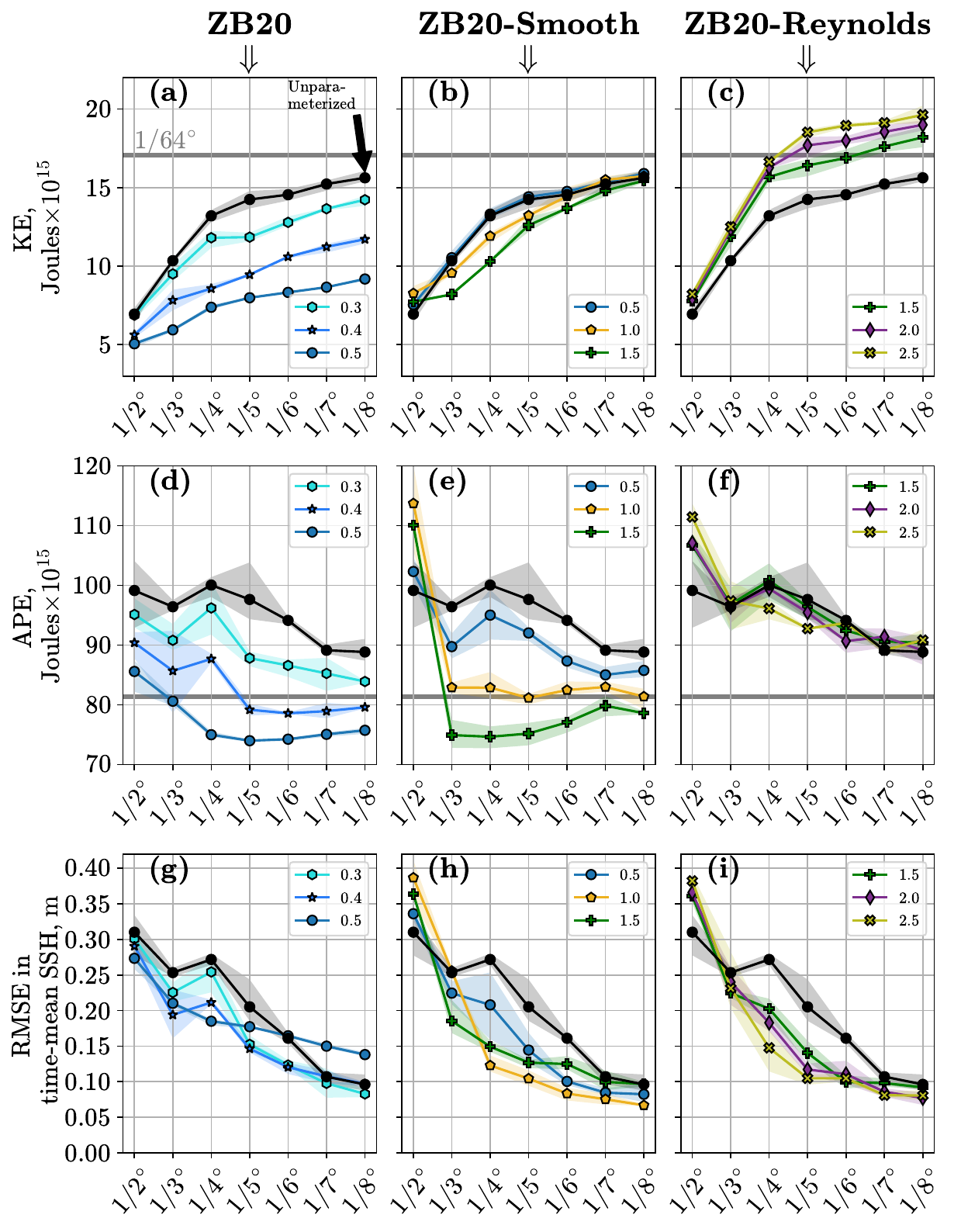}}
\caption{Sensitivity to the scaling coefficient $\gamma$ (shown in legend). Upper row: the time-mean kinetic energy (KE), middle row: the time-mean available potential energy (APE),  lower row: root mean squared error (RMSE) in the time-mean sea surface height (SSH). Left column shows unfiltered ZB20 parameterization and two rightmost columns show filtered ZB20 models. The black line shows the unparameterized model (biharmonic Smagorinsky). The gray horizontal line shows the $1/64^\circ$ model. The shading shows confidence intervals (see text).
}
\label{fig:ZB_models_generalization}
\end{figure} 

The parameterization effect of mesoscale eddies, which are partially resolved, should diminish as the grid is refined \cite{haidvogel2017numerical}. 
We achieve this property by informing the parameterization with the local grid spacing using Eq. \eqref{eq:kappa_bc}. Parameterizations with such scaling of the free coefficient are often referred to as scale-aware \cite{bachman2017scale,pearson2017evaluation}. To quantify the effect of the scale-aware parameter, we test the parameterizations in the Doube Gyre configuration in a range of seven simulations differing in grid resolution.

\subsection{Sensitivity to the scaling coefficient}
We perform experiments with the ZB20-based parameterizations for a range of resolutions from $1/2^\circ$ to $1/8^\circ$ and consider the sensitivity to the scaling coefficient $\gamma$. The effect of the subgrid parameterizations is quantified through the following metrics: the time-mean KE and APE, and the RMSE in the time-mean SSH. We compute the RMSE with respect to the 10-year averaged output of the coarse-grained high-resolution model $1/64^\circ$. We split the last 15 years of the simulation of the coarse resolution models into three 5-year segments. We compute the mean over the segments with confidence intervals provided by the min/max values over the different segments for every metric.

The impact of the ZB20-based parameterizations on all metrics is consistent with that described in Section \ref{sec:comparison_models} for most resolutions (Figure \ref{fig:ZB_models_generalization}). 
In particular, the ZB20 and ZB20-Smooth parameterizations spuriously reduce the KE (Figure \ref{fig:ZB_models_generalization}(a,b)) but improve the APE (Figure \ref{fig:ZB_models_generalization}(d,e)). Additionally, the spurious effect on the KE is smaller for the ZB20-Smooth parameterization than for the unfiltered ZB20. 
The ZB20-Reynolds model leads to an efficient backscatter parameterization: it increases the KE in the simulation (Figure \ref{fig:ZB_models_generalization}(c)) with little impact on the APE (Figure \ref{fig:ZB_models_generalization}(f)). All three subgrid parameterizations reduce the bias in SSH, with the simulation using the ZB20-Smooth parameterization having the lowest error and the simulation with the unfiltered ZB20 parameterization having the highest error (Figure \ref{fig:ZB_models_generalization}, lower row). 

The impact on the kinetic and potential energy for all three subgrid parameterizations is proportional to the scaling coefficient (Figure \ref{fig:ZB_models_generalization}, upper and middle rows). The default scaling coefficients used in the previous section  (ZB20-Smooth: $\gamma=1$, ZB20-Reynolds: $\gamma=2$) correspond to a compromise in reproducing the presented metrics for a range of resolutions. These non-dimensional coefficients can be kept constant without retuning, thus demonstrating an advantage of scale-aware tuning of the free parameter of the filtered ZB20 parameterizations. An overshoot in some metrics at $1/2^\circ$ resolution (Figure \ref{fig:ZB_models_generalization}(e)) occurs because of the coarse resolution (eddies are not permitted). In \ref{sec:additional_sensitivity}, we discuss the sensitivity to the number of filtering passes ($N$) and to the Smagorinsky coefficient ($C_S$).

\subsection{Comparison to the baseline parameterizations}

\begin{figure}[h!]
\centerline{\includegraphics[width=1.0\textwidth]{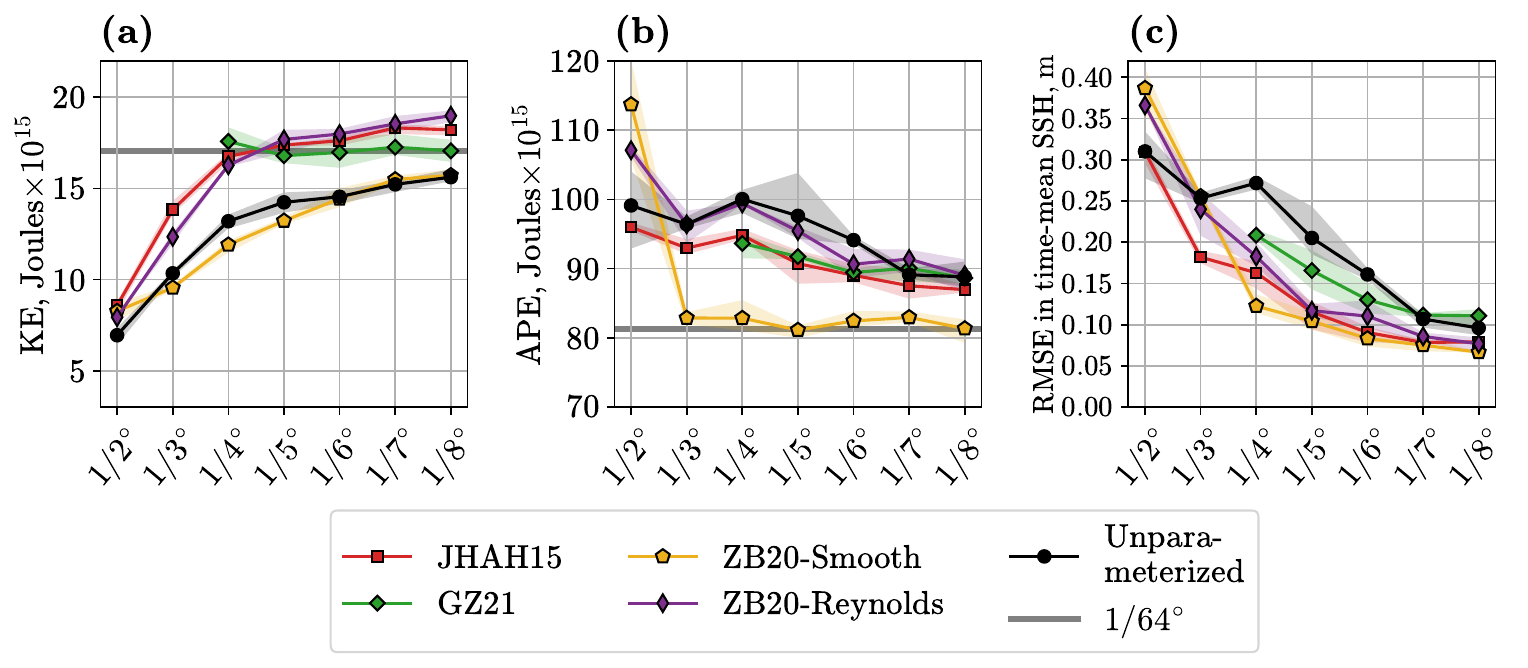}}
\caption{Similar to Figure \ref{fig:ZB_models_generalization}, but comparison of the filtered ZB20 parameterizations to the baselines of \citeA{jansen2015energy} and \citeA{guillaumin2021stochastic}. The scaling coefficient $\gamma$ is set to default: $1.0$ for ZB20-Smooth and $2.0$ for ZB20-Reynolds.}
\label{fig:generalization_JH_GZ}
\end{figure}

We compare the filtered subgrid parameterizations (ZB20-Smooth and ZB20-Reynolds) with the default parameter $\gamma$ to the baseline subgrid parameterizations of JHAH15 and GZ21. The subgrid parameterizations ZB20-Reynolds, GZ21, and JHAH15 are equally efficient in energizing the resolved flow, i.e., they parameterize the KE backscatter (Figure \ref{fig:generalization_JH_GZ}(a)) but have similar drawbacks. At the lowest resolutions ($1/2^\circ-1/3^\circ$), the ZB20-Reynolds and JHAH15 parameterizations underestimate the KE. At the highest resolutions ($1/5^\circ-1/8^\circ$), the ZB20-Reynolds and JHAH15 parameterizations slightly overestimate the KE without retuning; note that the GZ21 parameterization was tuned at every resolution to reproduce the KE exactly \cite{zhang2023implementation}. These three backscattering parameterizations have only a small impact on the APE (Figure \ref{fig:generalization_JH_GZ}(b)). The ZB20-Smooth parameterization demonstrates the best representation of the APE for a range of resolutions $1/3^\circ-1/8^\circ$ (Figure \ref{fig:generalization_JH_GZ}(b)). The SSH error with ZB20-Smooth is most evidently improved compared to all other parameterizations only at resolution $1/4^\circ$ discussed in Section \ref{sec:comparison_models}. The ZB20-Smooth leads to the most accurate representation of SSH at resolutions $1/5-1/8^\circ$, but its effect is comparable to the baseline simulation with JHAH15 parameterization. The ZB20-Smooth parameterization fails at resolutions coarser than $1/4^\circ$ (Figure \ref{fig:generalization_JH_GZ}(c)).

\section{Experiments in the NeverWorld2 (NW2) configuration} \label{sec:NW2}
In this section, we analyze the impact of the different subgrid parameterizations in a more complex adiabatic configuration of MOM6  -- the NeverWorld2 \cite<NW2,>{marques2022neverworld2} setup. This configuration spans the latitudes from $70^{\circ} \mathrm{S}$ to $70^{\circ} \mathrm{N}$, including the equator, with topography mimicking an idealized Mid-Atlantic ridge and a Drake Passage. The number of fluid layers is 15. This configuration was designed specifically to test mesoscale eddy parameterizations. Compared to the Double Gyre configuration, the NW2 setup has a stronger need for improving the energetics: the KE significantly depends on the resolution and increases by a factor of 4 when resolution is increased from $1/4^\circ$ to $1/32^\circ$ \cite{marques2022neverworld2}. Therefore, in this section, we target the $1/4^\circ$ resolution simulation to test the ZB20-based parameterizations and compare them to backscatter baseline parameterizations already implemented in NW2 (\citeA{yankovsky2023}, hereafter YBSZ24). 

\subsection{Numerical issues and tuning}

Preliminary experiments with the proposed ZB20-based parameterizations demonstrated numerical instabilities accompanied by various runtime errors, including too-large ocean velocities, interface height droping below the bathymetry, and NaN values in the prognostic fields. These issues are partly tied to the structure of isopycnals in NW2; for example, the significant part of the model area (40\%) is characterized by isopycnal layers with small thicknesses ($h_k \approx 10^{-2} \mathrm{m}$).  
To alleviate these numerical issues, we implemented two strategies:
\begin{itemize}
    \item Switch the discretization of the ZB20-based parameterizations from the energy-conserving form (Eq. \eqref{eq:ZB_scheme}) used in the Double Gyre to a non-conserving one (Eq. \eqref{eq:ZB_simple_scheme}). The non-conservative numerical scheme presumably introduces fewer aliasing errors because the multiplication operation follows the interpolation.
    \item Attenuate the parameterization in regions of geostrophically unbalanced flows. Similar to \citeA{klower2018energy, juricke2019ocean}, we introduce the following attenuation function:
    \begin{equation}
 \left(1+\frac{\sqrt{D^2+\widetilde{D}^2+\zeta^2}}{|f|} \right)^{-1}. \label{eq:Klower_attenuation}
    \end{equation}
    The ZB20 stress tensor is multiplied by the attenuation function before computing its divergence.
    The inclusion of $\zeta^2$ into Eq. \eqref{eq:Klower_attenuation} is a proposed modification, which was found to improve the numerical stability. The expression $D^2+\widetilde{D}^2+\zeta^2$ is proportional to the isotropic stress of the ZB20 parameterization, and thus the attenuation bounds the predicted momentum flux. 
    
    Note that the attenuation was inspired by a similar technique of shutting off the backscatter in high-strain regions proposed by YBSZ24. The difference between the two approaches is small: we attenuate the parameterization smoothly while they shut off the backscatter abruptly.  We use the inverse Coriolis parameter as a threshold time scale in the attenuation function while they use the time step (which is a non-physical parameter but useful because it is embedded in the numerical stability criteria). 
\end{itemize}

We further note that we have modified the Smagorinsky coefficient in the $1/4^\circ$ resolution model, compared to the original NW2 setup. We are using the commonly used (and our default value) of $C_S=0.06$ instead of the relatively large value of $C_S=0.2$ in \citeA{marques2022neverworld2}.
This change does not affect the reference unparametrized coarse resolution simulation (not shown). 

We decide to evaluate the parameterizations (ZB20-based and our baselines) by tuning them to roughly reproduce the total KE of the filtered NW2 high-resolution simulation while minimizing the difference across tuning coefficients. 
In our preliminary experiments, we noted that the unfiltered ZB20 parameterization could not be tuned to reproduce the KE of the filtered high-resolution model. The increase in the KE compared to the unparameterized simulation was never larger than $\approx 25\%$, with only small improvements in the mean state, thus we do not show results for the unfiltered ZB20 parameterization in this section. The ZB20-Smooth and ZB20-Reynolds parameterizations were tuned by setting the scaling coefficient to $\gamma=2.5$. 

To summarize, compared to the Double Gyre case, we changed the discretization scheme of the ZB20 parameterization, introduced the attenuation function, and increased the scaling coefficient. 

\subsection{Results}
\begin{figure}[h!]
\centerline{\includegraphics[width=1.0\textwidth]{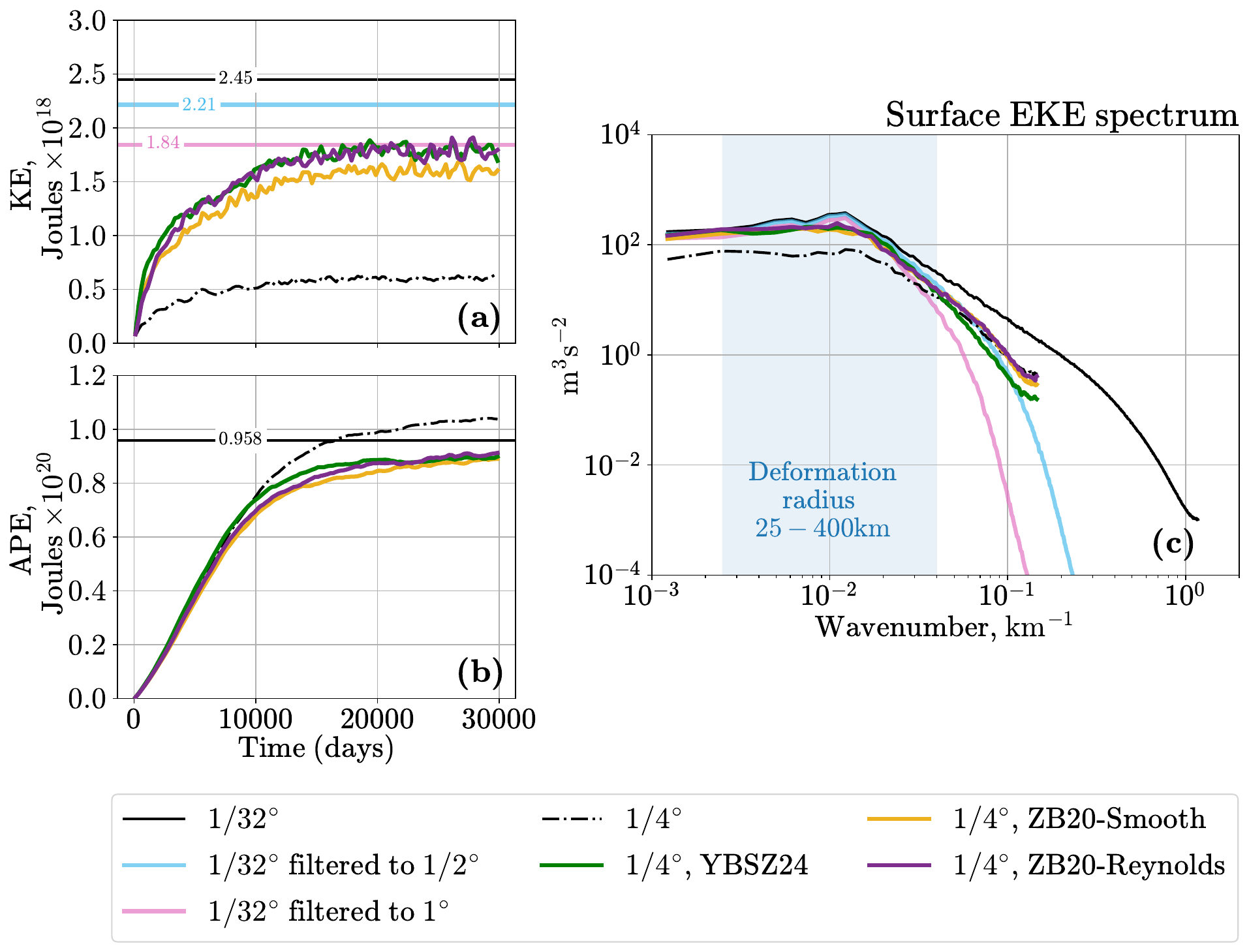}}
\caption{NeverWorld2 configuration. Time series of the  (a) kinetic energy (KE) and (b) available potential energy (APE). The time series in $1/4^\circ$ runs are smoothed in time with a window size of 250 days, while in $1/32^\circ$ run, we provide average values over the last 100 days. 
(c) The EKE zonal spectrum at the surface averaged over 100 days and over latitudes. 
}
\label{fig:NW2_series_spectrum}
\end{figure}

\begin{figure}[h!]
\centerline{\includegraphics[width=1.0\textwidth]{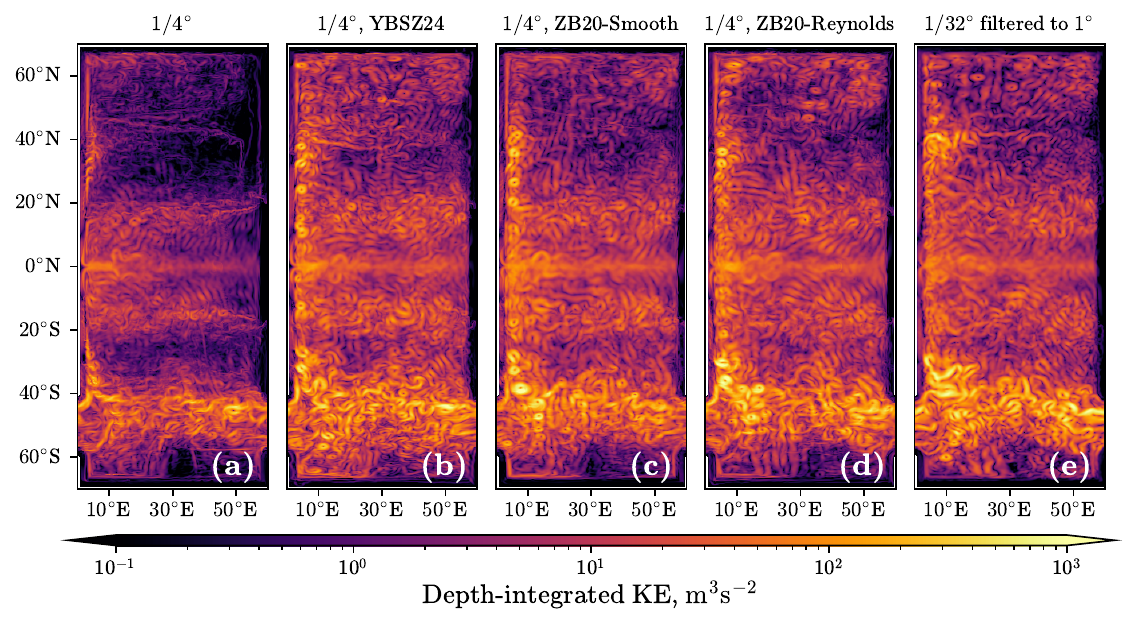}}
\caption{Snapshot of the depth-integrated kinetic energy. The coarse models at resolution $1/4^\circ$: (a) unparameterized model (biharmonic Smagorinsky), (b) \citeA{yankovsky2023} (YBSZ24), (c) and (d) filtered ZB20 parameterizations. In panel (e), we show the filtered high-resolution model $1/32^\circ$ with filter width $1^\circ$.}
\label{fig:NW2_snapshot}
\end{figure}

\begin{figure}[h!]
\centerline{\includegraphics[width=1.0\textwidth]{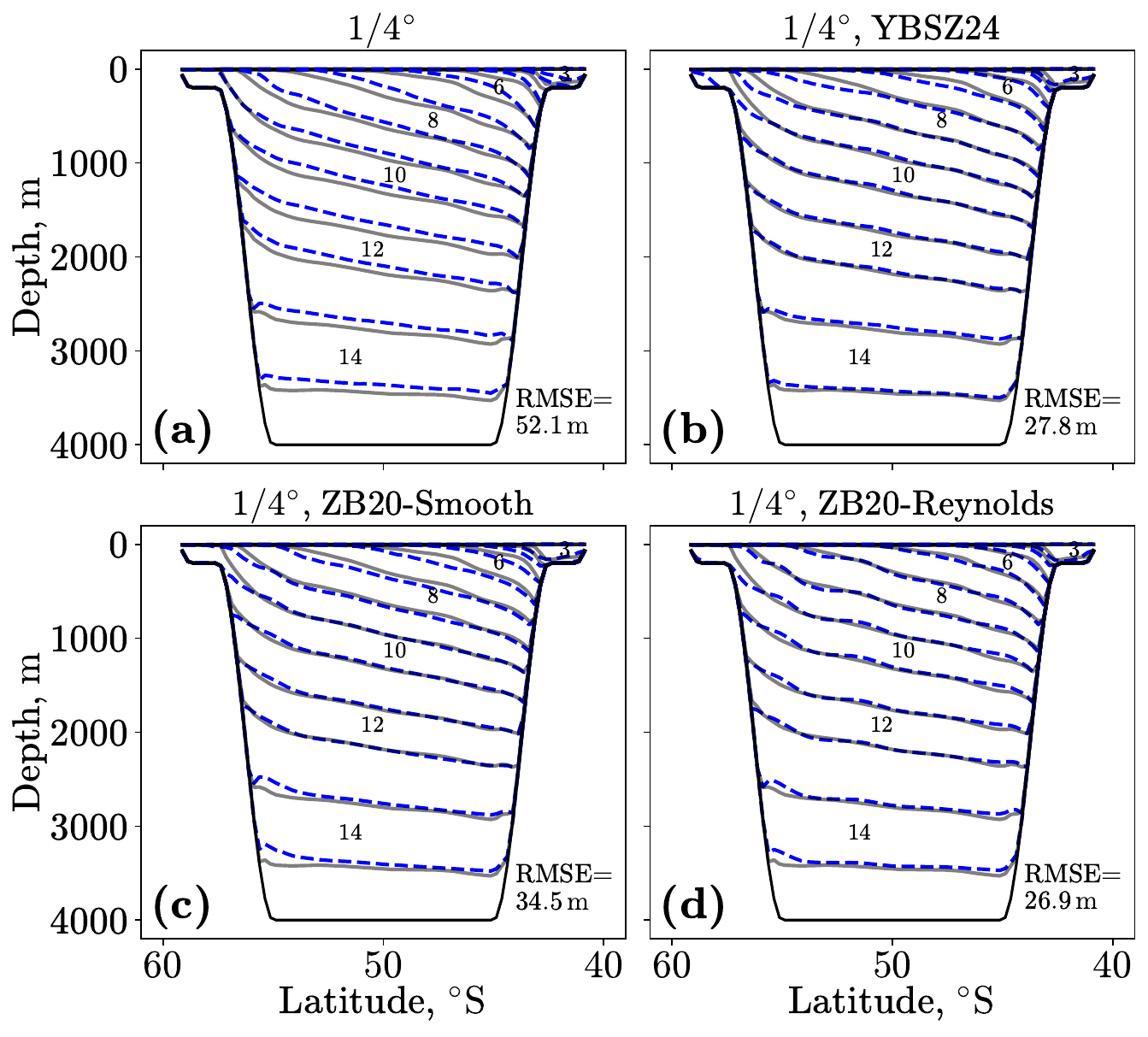}}
\caption{The time-mean interfaces in the meridional transect of Drake Passage (Longitude $0^{\circ}\mathrm{E}$) averaged over 1000 days. The blue dashed lines show the experiments at resolution $1/4^\circ$: (a) unparameterized model, (b) \citeA{yankovsky2023} (YBSZ24), (c) and (d) filtered ZB20 parameterizations. Gray lines show the interfaces of the high-resolution model $1/32^\circ$. The layer numbers (equal to 14, 12,...) are provided.}
\label{fig:NW2_Drake}
\end{figure} 

The high-resolution simulation $1/32^\circ$ was spun up in multiple stages \cite{marques2022neverworld2}. We use snapshots from the last 100 days and time-mean fields over the last 1000 days for analysis. 
We ran the coarse parameterized and unparameterized models at resolution $1/4^\circ$ for 30000 days, starting from rest. 
To compare the coarse and high-resolution simulations, we filter the output of the  $1/32^\circ$ high-resolution simulations with filter widths $1^\circ$ and $1/2^\circ$. That is, we assume that the effective resolution of the coarse resolution models is lower than the nominal resolution given by the coarse grid \cite{skamarock2004evaluating, soufflet2016effective}.

The baseline GZ21 parameterization, used in the Double Gyre experiment, struggles to generalize to the NW2 configuration, and thus, we omit the results. 
Rather than use the JHAH15 parameterization, as for the Double Gyre configuration, we used the most up-to-date version of the JHAH15 parameterization from YBSZ24, which was specifically tuned to represent mesoscale eddies in the NW2 configuration. 

The YBSZ24 parameterization attributes a vertical structure to the anti-viscosity coefficient, tuned to match the energetics of the NW2 simulation in their original study.  
This led to a choice of vertical structure that follows the equivalent barotropic mode raised to the 2nd power. 
Similarly to the ZB20-based parameterizations, the coarse resolution simulation using the YBSZ24 parameterization was tuned to match the kinetic energy of the filtered high-resolution run, using an anti-viscosity scaling coefficient of $-0.3$. 
Note, however, that the YBSZ24 parameterization in NW2 requires a much larger Smagorinsky coefficient ($C_S=0.2$) to remain stable compared to the unparameterized and ZB20-based parameterized models ($C_S=0.06$). 

We compare the coarse models to the high-resolution simulation (Fig.~\ref{fig:NW2_series_spectrum}, black solid line) and its filtered versions (Fig.~\ref{fig:NW2_series_spectrum}, light blue and pink lines). 
We do not show the APE of the filtered high-resolution model because it is significantly affected by the implementation of the filtering algorithm near the ocean bottom.

The unparameterized model ($1/4^\circ$) has a KE that is approximately 4 times lower than the KE of the $1/32^\circ$ model and 3 times lower than the KE of the filtered $1/32^\circ$ model (Figure \ref{fig:NW2_series_spectrum}(a)).  
The APE of the unparameterized model is too high compared to the high-resolution simulation (Figure \ref{fig:NW2_series_spectrum}(b)) since the coarse resolution model underestimates the barotropization of the flow \cite{kjellsson2017impact}. 
The ZB20-Smooth and ZB20-Reynolds parameterizations increase the KE approximately 3 times and reduce the APE below the APE of the high-resolution simulation (as we expect for the filtered solution). The ZB20-Smooth parameterization is slightly more efficient in reducing the APE, and the ZB20-Reynolds is slightly more efficient in increasing the KE (Figure \ref{fig:NW2_series_spectrum}(a,b)) for the same scaling coefficient $\gamma$.
The YBSZ24 parameterization is equally efficient in reducing APE and increasing KE. The equilibration of the parametrized run with the YBSZ24 parameterization is faster than the parametrized simulations with the filtered ZB20 closures.

Figure \ref{fig:NW2_series_spectrum}(c) shows the zonal EKE spectrum at the surface, $1/2\left(|\mathcal{F}(u')|^2 + |\mathcal{F}(v')|^2\right)$, averaged in time over 100 days and over all latitudes. The Fourier transform $\mathcal{F}$ is computed in the zonal direction with Hann window and linear detrending. The eddy velocities $u'$ and $v'$ are defined as a deviation from the $1000$-day mean velocities at the surface. The filtered ZB20 parameterizations and the YBSZ24 parameterization increase the EKE spectrum density at large scales (Figure \ref{fig:NW2_series_spectrum}(c)). 
Considering the shape of the EKE spectrum at small scales, the filtered ZB20 models are closest to the $1/32^\circ$ model filtered with the filter width $1/2^\circ$ (Figure \ref{fig:NW2_series_spectrum}(c)). However, the total kinetic energy better matches the filter scale of $1^\circ$ (Figure \ref{fig:NW2_series_spectrum}(a)). A further increase in KE is possible but at the expense of the APE presumably becoming unphysical. These findings reveal the difficulties in determining the effective resolution of the parameterized simulations.
The simulation with the YBSZ24 parameterization possesses less energy at small scales at the surface, likely due to the high value of the Smagorinsky coefficient, which is necessary for numerical stability (in addition to the local shut-off described in their original study). 


The snapshots of the depth-integrated KE are shown in Figure \ref{fig:NW2_snapshot}. The filtered ZB20 models considerably energize the eddies in all parts of the domain compared to the unparameterized model and in accordance with the high-resolution simulation. The simulation with the YBSZ24 parameterization is also successful at increasing the KE throughout the domain, with only small differences compared to the ZB20-based parameterized simulations. The major difference between the coarse parameterized models and the high-resolution simulation is the presence of coherent eddies near the western boundary (Figure \ref{fig:NW2_snapshot}(b,c,d)), including in the YBSZ24 parametrized simulation. Presumably, the attenuation function (Eq. \eqref{eq:Klower_attenuation}) was needed to bound the growth of these and similar eddies in the filtered ZB20 models. However, tracking how a particular eddy growth contributes to the development of numerical instabilities is difficult.

\begin{table}
	\begin{center}
		\begin{tabular}{l|cccc}
			         & $0^{\circ}\mathrm{E}$ ($\mathrm{m}$)  & $30^{\circ}\mathrm{E}$ ($\mathrm{m}$) & $45^{\circ}\mathrm{E}$ ($\mathrm{m}$) & SSH ($\mathrm{m}$)\\ 
            \hline 
			$1/4^{\circ}$     	            &  52.1  & 40.3 & 34.6 & 0.101       \\
			$1/4^{\circ}$, YBSZ24           &  27.8  & 20.7 & 21.3 & \textbf{0.071}     \\
			$1/4^{\circ}$, ZB20-Smooth      &  34.5  & 20.8 & 24.9 & 0.090       \\
                $1/4^{\circ}$, ZB20-Reynolds    &  \textbf{26.9}  & \textbf{18.4} & \textbf{18.5} & 0.080
            
		\end{tabular}
		\caption{The root mean squared errors (RMSE) in 1000-day averaged position of interfaces over three meridional transects at ($0^{\circ}\mathrm{E}$, $60^\circ \mathrm{S}$-$40^\circ \mathrm{S}$) (Drake Passage, also shown in Figure \ref{fig:NW2_Drake}), ($30^{\circ}\mathrm{E}$, $70^\circ \mathrm{S}$-$70^\circ \mathrm{N}$) (Mid-Atlantic ridge) and ($45^{\circ}\mathrm{E}$, $70^\circ \mathrm{S}$-$70^\circ \mathrm{N}$) (similar to \citeA{yankovsky2023}). Also, we provide RMSE in 1000-day averaged sea surface height (SSH). The error is computed with respect to $1/32^{\circ}$ model.
		}
		\label{tab:RMSE_NW2}
	\end{center}
\end{table}

The largest portion of the APE of the mean state is described by the vertical structure of the isopycnal interfaces in the Drake Passage ($60^{\circ}\mathrm{S}$-$40^{\circ}\mathrm{S}$), shown in Figure \ref{fig:NW2_Drake} at longitude $0^{\circ} \mathrm{E}$. The mesoscale eddies extract potential energy from the mean state and act to flatten the isopycnals. The coarse unparameterized model poorly resolves mesoscale eddies, and consequently, its isopycnals are too steep (Figure \ref{fig:NW2_Drake}(a)). Both filtered ZB20 parameterizations and the YBSZ24 parameterization result in reducing the potential energy and better reproducing the isopycnal structure in Drake Passage (Figure \ref{fig:NW2_Drake}(b,c,d)). In the same figure, we provide the RMSE values with respect to the high-resolution simulation. 
The biggest improvement is achieved with the ZB20-Reynolds parameterization, which has the lowest RMSE. The improvements are visible in the upper ocean in particular, compared to the ZB20-Smooth and the YBSZ24 parameterizations. 
However, the bottom two layers are better represented in the YBSZ24 parameterized run than in the ZB20-based simulations. 
The performance of the filtered ZB20 parameterizations is confirmed according to additional metrics quantifying the mean state. These metrics include the RMSE in the vertical structure of isopycnals in three transects; see Table \ref{tab:RMSE_NW2}. 
All parametrized simulations reduce biases, with the ZB20-Reynolds parameterized simulation showing the largest improvement using these metrics.
When considering the RMSE in the time-mean SSH (Table \ref{tab:RMSE_NW2}), all parametrized models decrease the mean bias, with the YBSZ24 parameterization performing best. However, all improvements in the mean state are rather small according to this metric. 



\section{Conclusions and discussion}

In this work, we implemented the data-driven mesoscale eddy parameterization introduced by \citeA{zanna2020data} (ZB20) into the GFDL ocean model, MOM6, and tested it in two idealized configurations: the Double Gyre and NeverWorld2 (NW2). 
The ZB20 parameterization, machine-learned from data,  predicts the subgrid momentum fluxes and, in particular, captures the kinetic energy (KE) backscatter, that is, the inverse energy cascade from the subgrid to resolved scales. 
Our main findings are as follows:
\begin{itemize}
    \item 
    The original parameterization was found to generate numerical noise near the grid scale. We propose two filtering schemes to reduce the generation of numerical noise and isolate the large-scale backscatter effect of the parameterization. The first scheme (ZB20-Smooth) applies a low-pass filter to the ZB20 stress tensor, and the second scheme (ZB20-Reynolds) additionally applies a high-pass filter to the velocity gradients \cite{perezhogin2023subgrid}.
    \item The free parameter of the ZB20-based parameterizations is scale-aware. Here, we show that the parameterizations can be used for a range of eddy-permitting resolutions without retuning.
    \item
    In the Double Gyre configuration, the ZB20-Reynolds parameterization effectively energizes the resolved flow and performs similarly to the \citeA{jansen2015energy} (JHAH15) and \citeA{guillaumin2021stochastic} (GZ21) backscatter parameterizations at eddy-permitting resolutions ($1/4^\circ-1/8^\circ$). In this configuration, the ZB20-Smooth is less efficient in parameterizing the backscatter but outperforms other parameterizations in the SSH and potential energy predictions at resolutions $1/4^\circ-1/8^\circ$. All subgrid parameterizations perform poorly at the coarsest resolutions ($1/2^\circ-1/3^\circ$), as expected.
    \item In the NW2 configuration, the filtered ZB20 parameterizations improve the model energetics, namely kinetic and potential energy reservoirs, and the energy power spectrum. The ZB20-based parameterizations improve aspects of the climatological mean state (here the vertical structure of the isopycnals and the SSH). The ZB20-based parameterizations perform better or as well as the anti-viscosity parameterization of \citeA{yankovsky2023}, which imposes a vertical structure to the anti-viscosity coefficient.  The ZB20-based parameterization does not need to impose a vertical structure to its coefficient, the 3D flow dependence is encapsulated in the stress tensor itself.  
    \item  The direct effect of mesoscale eddy parameterizations in improving the mean state depends on the ocean configuration. In the NW2 configuration, the subgrid parameterization increases the KE of the resolved flow by a factor of 3. This significantly enhances the effect by resolved eddies which act to flatten isopycnals and reduce the potential energy. In the Double Gyre configuration, on the other hand, the mean state can be improved without energizing the resolved eddies but instead by imposing strong mean subgrid stress \cite{kjellsson2017impact}.
\end{itemize}


Our methodology enables researchers and ocean modelers to implement and test subgrid machine-learning parameterizations in state-of-the-art ocean models.
As with all parameterizations in current climate models, the ZB20 parameterization required tuning. The succinct, interpretable form of the subgrid ZB20 model allowed us to study its physical and numerical properties in detail.  
We leveraged the filtering schemes to extract the parameterization effect on the large-scale flow and to avoid the grid-scale numerical issues. By testing various discretizations of the subgrid model, we were able to find a numerically stable scheme for online simulations.  Our filtering schemes can be potentially applied to improve the performance of our baseline parameterizations (JHAH15, BSZ24 and GZ21); though such testing is beyond the scope of this study.  We note that filtering schemes have been consistently used in the literature, showing a potential to improve the performance of other mesoscale eddy parameterizations \cite{grooms2015numerical, juricke2020ocean, mak2023scale, grooms2023backscatter, bagaeva2023}. 

While applying the filtering schemes allows us to improve the physical and numerical properties of the ZB20 parameterization, several challenges remain. (1) It is difficult to find a single filtering scheme that simultaneously improves the mean state and kinetic/potential energies. (2) The proposed filtering approach is applied and tuned a posteriori; how to learn it directly from data is an open question. (3) The filtered ZB20 parameterizations do not impact the grid scale flow, therefore a supplementary subgrid dissipative model is needed (in this work, we utilize a biharmonic Smagorinsky model). (4) Without filters, the ZB20 parameterization is computationally cheap ($2.5\%$ of total runtime). However, the filtered parameterizations can take $4-6\%$ of total runtime.

Possible future improvements to the proposed parameterization include: (1) Coupling the ZB20-based parameterizations with the subgrid kinetic energy equation, as in energetically-consistent parameterizations \cite{jansen2015energy, mak2018implementation}.  (2) Including spatial non-locality \cite{wang2022non} or temporal memory \cite{zanna2017scale} during the training process to potentially reduce the need for a posteriori tuning. (3) Using a neural network to improve the prediction of the subgrid stress from the same input features. 
(4) Informing the subgrid model with local physical parameters to improve generalization to unseen flow regimes \cite{hallberg2013using, bachman2017scale, jansen2019toward}. 

An additional interesting future direction is to apply the developed subgrid parameterizations in eddy-permitting global ocean models to attempt addressing long-standing biases such as the North Atlantic cold bias \cite{wang2014global, flato2014evaluation, chang2023remote}.

\appendix
\section{Curvilinear coordinates, varying layer thickness, numerical schemes and boundary conditions} \label{appendix:model_formulation}
We modify the original parameterization (Eqs. \eqref{eq:ZB_momentum_forcing}, \eqref{eq:ZB_momentum_flux}) to account for curvilinear coordinates and varying layer thickness. Also, we propose a numerical discretization scheme and boundary conditions.

\subsection{Computation of the stress tensor}
The components of the stress tensor $\mathbf{T}$ depend uniquely on the gradients of the velocity field ($D$, $\widetilde{D}$, $\zeta$).  The computation of these gradients depends on the coordinate system. In generalized curvilinear orthogonal coordinates the expressions for $D$ and $\widetilde{D}$ must be changed to (Appendix A.b in \citeA{griffies2000biharmonic} and Section 17.10.2 in \citeA{griffies2018fundamentals}):
\begin{gather}
    D=\Delta_y \partial_x (v/\Delta_y) + \Delta_x\partial_y (u/\Delta_x), \\
    \widetilde{D}=\Delta_y \partial_x (u/ \Delta_y) - \Delta_x \partial_y (v/\Delta_x),
\end{gather}
where $\Delta_x$ and $\Delta_y$ are local grid spacings that are proportional to the Lame coefficients. 

Following a similar approach, we compute the relative vorticity as follows:
\begin{equation}
    \zeta=\Delta_y \partial_x (v/\Delta_y)-\Delta_x \partial_y (u/\Delta_x). \label{eq:rel_vort}
\end{equation}
Note that the relative vorticity can be alternatively computed using the contour integral divided by area, that is  $\zeta = \Delta_y^{-1}\partial_x (\Delta_y v)-\Delta_x^{-1}\partial_y(\Delta_x u)$, see Section 2.3.2 in \citeA{madec2008nemo}. We found that both approaches give close results and use Eq. \eqref{eq:rel_vort} for simplicity.

\subsection{Divergence of momentum flux}
Following previous work on viscous operators in ocean models (Eqs. (A3) and (A4) in  \citeA{griffies2000biharmonic} and Section 17.10.3 in \citeA{griffies2018fundamentals}), we modify the divergence of the stress tensor (Eq. \eqref{eq:ZB_momentum_forcing}):
\begin{equation}
    \mathbf{S} = 
    \frac{1}{h} \nabla \cdot (h \mathbf{T}) = \frac{1}{h}
    \begin{pmatrix}
        \frac{1}{\Delta_y^2}\partial_x (\Delta_y^2 h T_{xx}) + \frac{1}{\Delta_x^2}\partial_y (\Delta_x^2 h T_{xy}) \\
        \frac{1}{\Delta_y^2} \partial_x (\Delta_y^2 h T_{xy}) + \frac{1}{\Delta_x^2}\partial_y (\Delta_x^2 h T_{yy})
    \end{pmatrix}, \label{eq:ZB_thickness}
\end{equation}
where we account for curvilinear coordinates with terms including $\Delta_x$ and $\Delta_y$ and for varying layer thickness with terms including $h$. The components of the subgrid forcing parameterization $\mathbf{S}$ have the dimensions of acceleration, i.e. $\mathrm{length}^1~\mathrm{time}^{-2}$. The components of the stress tensor ($T_{xx}, T_{xy},...$) have the dimensions of $\mathrm{length}^2~\mathrm{time}^{-2}$.

Accounting for varying thickness allows to build a parameterization that conserves the integral of momentum up to the boundary fluxes:
\begin{eqnarray}
     \partial_t \int h \mathbf{u} \, dxdy & = & \cdots + \int h \partial_t \mathbf{u} \, dxdy = \cdots + \int h \mathbf{S} \, dxdy = \cdots + \\
      && \cancelto{0}{\int \nabla \cdot (h \mathbf{T}) \, dxdy}.
\end{eqnarray}

\subsection{Numerical discretization} \label{appendix:discretization}
The rate of change of the KE due to the parameterization (Eq. \eqref{eq:ZB_thickness}) after integration by parts is given by:
\begin{eqnarray}
    \partial_t \int \frac{1}{2} h |\mathbf{u}|^2 \, dxdy & = & \cdots + \int (h \mathbf{u}) \mathbf{S} \, dxdy = \cdots + \int \mathbf{u} \cdot \nabla \cdot (h \mathbf{T}) \, dxdy  = \cdots \\ && - \int h \mathbf{T} : (\nabla \mathbf{u}) \, dxdy,
\end{eqnarray}
where $\nabla \mathbf{u}$ is the velocity gradient tensor and $(:)$ is the tensor contraction over two indices \cite{eyink1995local}. In Cartesian coordinates the components of velocity gradient tensor have a simple form $(\nabla \mathbf{u})_{ij} = \partial_j u_i$. The energy contribution from the deviatoric component of $\mathbf{T}$ (Eq. \eqref{eq:ZB_momentum_flux}) is zero because it is orthogonal to the velocity gradient tensor:
\begin{equation}
    2 \mathbf{T}_{\mathrm{d}} : (\nabla \mathbf{u}) = 
    \begin{bmatrix}
        - \zeta D & \zeta \widetilde{D} \\
        \zeta \widetilde{D} & \zeta D 
    \end{bmatrix}
    :
    \underbrace{\begin{bmatrix}
    \widetilde{D} & D \\
    D  &-\widetilde{D}
    \end{bmatrix}}_{\substack{\text{deviatoric} \\ \text{symmetric} \\ \text{part of $2\nabla \mathbf{u}$}}} = 
    - \zeta D \widetilde{D} + \zeta \widetilde{D} D + \zeta \widetilde{D} D - \zeta  D \widetilde{D} = 0. \label{eq:ZB_energy_flux}
\end{equation}

However, the Eq. \eqref{eq:ZB_energy_flux} does not hold numerically when the Arakawa C  grid staggering is used,  because $D$, $\zeta$ and $T_{xy}$ are defined in the corner of the grid cell, but $T_{xx}$, $T_{yy}$ and $\widetilde{D}$ are defined in the center of the grid cell. We propose the following energy-conserving discretization of the deviatoric stress $\mathbf{T}_{\mathrm{d}}$:
\begin{equation}
    \underbrace{
    \begin{bmatrix}
        - \overline{\zeta D}^{x,y} & \zeta \overline{\widetilde{D}}^{x,y} \\
        \zeta \overline{\widetilde{D}}^{x,y} & \overline{\zeta D}^{x,y} 
    \end{bmatrix}}_{\text{numerical scheme}}
    :
    \begin{bmatrix}
    \widetilde{D} & D \\
    D  &-\widetilde{D}
    \end{bmatrix} = - \underbrace{2 \overline{\zeta D}^{x,y}\widetilde{D}}_{\text{cell center}} + \underbrace{2 \zeta D \overline{\widetilde{D}}^{x,y}}_{\text{cell corner}}, \label{eq:ZB_scheme}
\end{equation}
where $\overline{(\cdot)}^{x,y}$ is a linear interpolation from corner to center or vice versa. Terms in the RHS of Eq. \eqref{eq:ZB_scheme} are defined in different points but cancel after summation over the domain because the interpolation operator is self-adjoint up to the boundary conditions (Section 4.1.2 in \citeA{madec2008nemo}). Energy-conserving discretization on a non-uniform grid is given by computing the interpolation $\overline{\zeta D}^{x,y}$ in a conservative way, i.e. by weighting with the local grid cell area. Note that exact energy conservation property is lost when coefficient $\kappa_{BC}$ or layer thickness $h$ vary spatially or when the spatial filters are applied.

For the isotropic part of $\mathbf{T}$ (Eq. \eqref{eq:ZB_momentum_flux}), we use the simplest numerical scheme because no properties are known:
\begin{equation}
    \mathbf{T}_{\mathrm{I}} = \left( (\overline{\zeta}^{x,y})^2 + (\overline{D}^{x,y})^2 + \widetilde{D}^2 \right)
    \begin{bmatrix}
        1 & 0 \\
        0 & 1
    \end{bmatrix}.
\end{equation}

The energy-conserving discretization (Eq. \eqref{eq:ZB_scheme}) is used in the Double Gyre configuration. In NW2 runs, we found that the following approximation of the deviatoric component of $\mathbf{T}$ demonstrates better numerical stability properties: 
\begin{equation}
    \mathbf{T}_{\mathrm{d}}=\begin{bmatrix}
        - \overline{\zeta}^{x,y} \overline{D}^{x,y} & \zeta \overline{\widetilde{D}}^{x,y} \\
        \zeta \overline{\widetilde{D}}^{x,y} & \overline{\zeta}^{x,y} \overline{D}^{x,y}
    \end{bmatrix}. \label{eq:ZB_simple_scheme}
\end{equation}

\subsection{Boundary conditions}
We apply an analog of the free-slip boundary condition: the momentum flux through the boundary is zero. On the Arakawa C grid, this is achieved by setting $T_{xy}=0$ on the boundary. Zero boundary conditions are also used for the filtering operations and interpolations on the staggered grid.

\section{Computation of kinetic and potential energy} \label{appendix:KE_APE}

The KE integrated over the fluid layers and horizontal coordinates in Joules is defined as:
\begin{equation}
    \mathrm{KE} = \frac{1}{2} \sum_k \int \rho_0 |\mathbf{u}_k|^2 h_k \, dx dy.
\end{equation}
The KE of the mean flow ($\overline{\mathbf{u}}^t_k$, $\overline{h}^t_k$) is referred to as a mean kinetic energy (MKE). The eddy kinetic energy (EKE) is defined as $\mathrm{EKE} = \overline{\mathrm{KE}}^t-\mathrm{MKE}$. The potential energy (PE) summed over interfaces in Joules is defined as:
\begin{equation}
    \mathrm{PE} = \frac{1}{2} \sum_k \int \rho_0 g_{k+1/2}' \eta_{k+1/2}^2 \, dxdy.
\end{equation}
The available potential energy (APE) is the PE minus the potential energy of the resting state given by $\eta_{k+1/2}^{\mathrm{ref}} = \max(z^0_{k+1/2}, -H(x,y))$, where $z^0_{k+1/2}$ is the constant nominal position of the interfaces and $H\geq0$ is the depth \cite{marques2022neverworld2}. Mean potential energy (MPE) is the APE of the mean flow $\overline{\eta}^t_{k+1/2}$, and eddy potential energy (EPE) is given by $\mathrm{EPE}=\overline{\mathrm{APE}}^t-\mathrm{MPE}$.

\section{Additional sensitivity studies} \label{sec:additional_sensitivity}
\begin{figure}[h!]
\centerline{\includegraphics[width=0.8\textwidth]{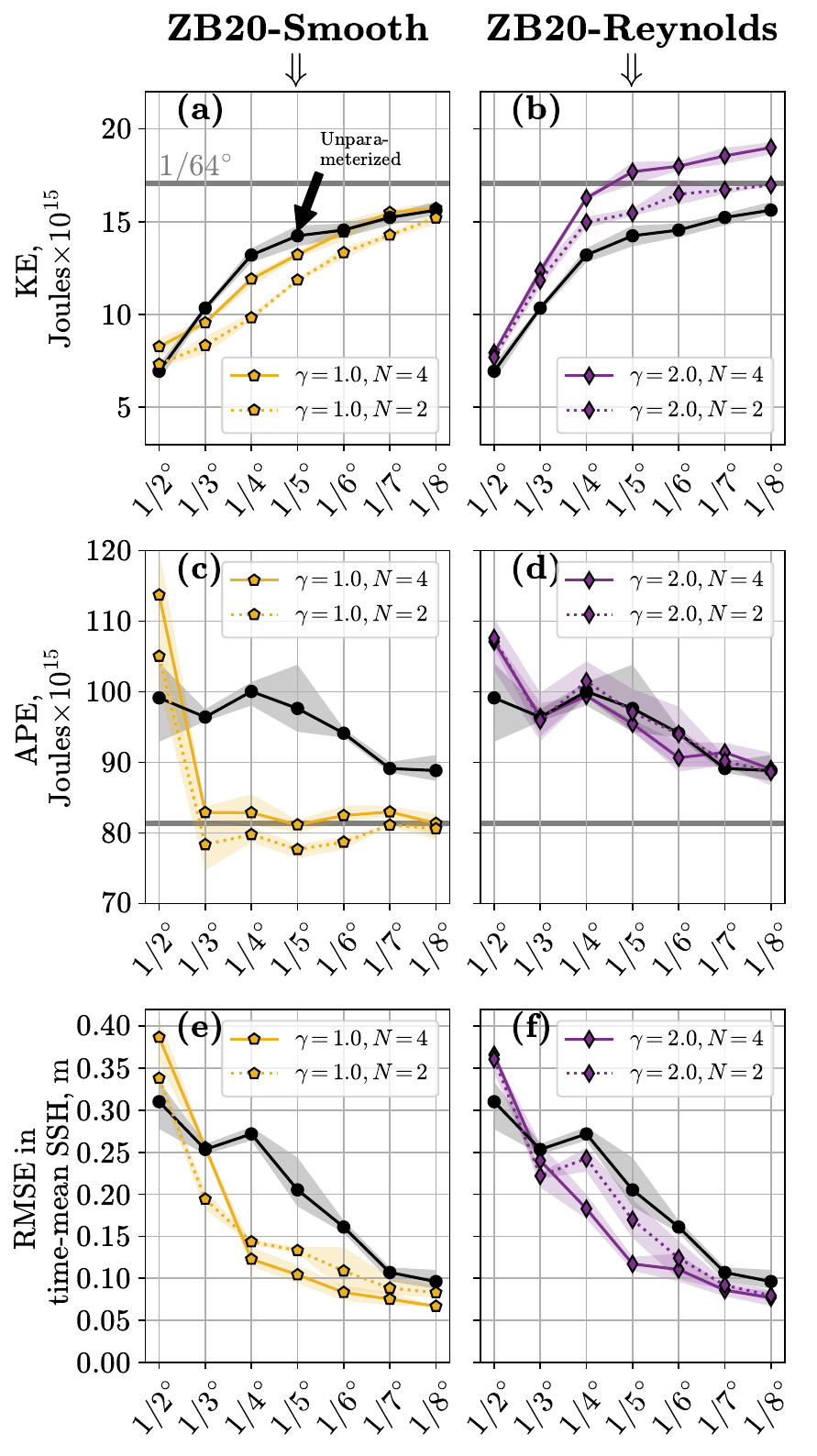}}
\caption{Similar to Figure \ref{fig:ZB_models_generalization}, but showing the sensitivity to the number of the filter passes $N$ for the filtered ZB20 parameterizations. The solid lines show the default value of the filter passes ($N=4$), and the dotted lines show a smaller value ($N=2$).}
\label{fig:sensitivity_N}
\end{figure}

In Figure \ref{fig:sensitivity_N} we show the sensitivity of the online metrics to the number of filter passes ($N$) for the ZB20-Smooth and ZB20-Reynolds parameterizations. For the ZB20-Smooth, the effect of reducing APE is slightly stronger for lower $N$ at the same scaling coefficient $\gamma$ (Figure \ref{fig:sensitivity_N}(b)). However, the undesirable impact on the KE is also stronger (Figure \ref{fig:sensitivity_N}(a)). Considering the error in SSH, the optimal number of filters is $N=2$ for coarse resolutions ($1/2^\circ-1/3^\circ$) and the optimal number is $N=4$ for higher resolutions (Figure \ref{fig:sensitivity_N}(c)). For the  ZB20-Reynolds parameterization, the impact on the KE is higher for higher $N$ (Figure \ref{fig:sensitivity_N}(d)). An impact on the SSH metric is also more significant for larger $N$ (Figure \ref{fig:sensitivity_N}(f)). Overall, a slightly more beneficial impact on energetic metrics (KE, APE) is observed in the case of a larger number of filters $N=4$. However, testing in different configurations is required to determine the optimal parameter $N$.

\begin{figure}[h!]
\centerline{\includegraphics[width=1.0\textwidth]{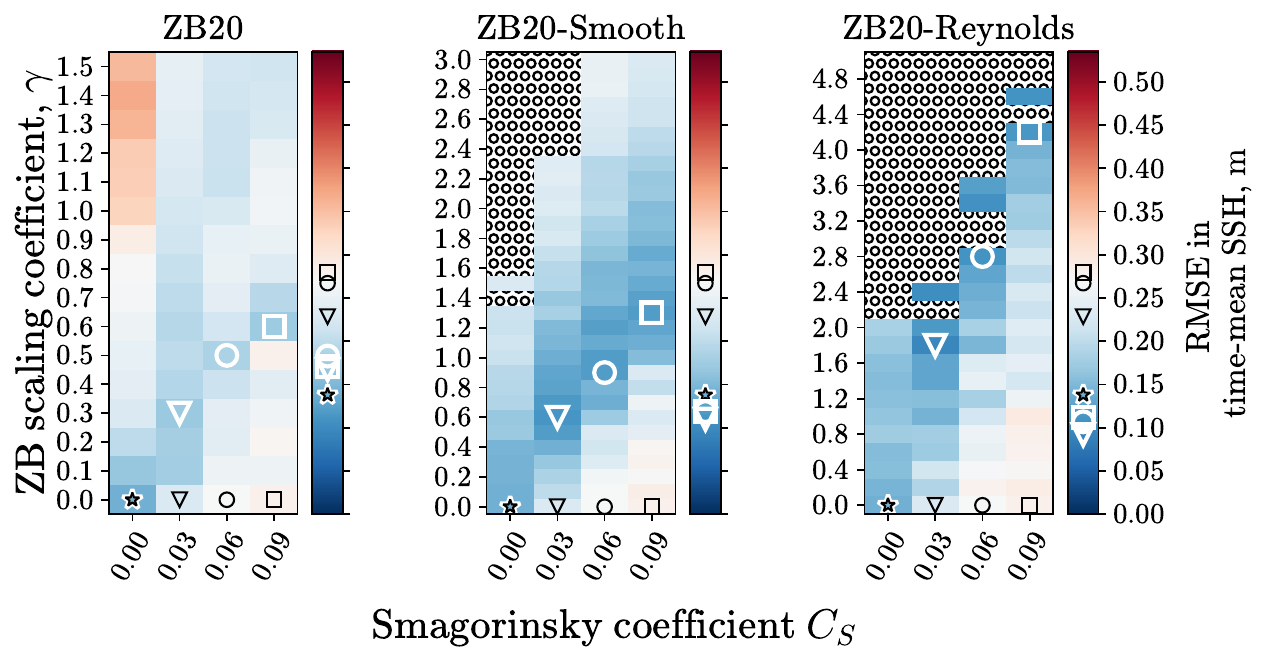}}
\caption{The sensitivity to the Smagorinsky coefficient $C_S$ and scaling coefficient $\gamma$ in the ZB20-based parameterizations. The considered metric is the root mean squared error (RMSE) in the representation of the time-mean sea surface height (SSH). The ocean grid resolution is $1/4^\circ$.  Every small box shows a single simulation. The hatch area shows unstable simulations.}
\label{fig:sensitivity_Cs}
\end{figure}

In Figure \ref{fig:sensitivity_Cs}, we show the sensitivity to the Smagorinsky coefficient. We consider the bias in SSH prediction because it is sensitive to including any of the parameterizations (ZB20, ZB20-Smooth, ZB20-Reynolds). The unparameterized models with different Smagorinsky coefficients are shown in black markers. White markers show the optimal scaling coefficient $\gamma$ for a given Smagorinsky constant. Note that we include inviscid simulations ($C_S=0.00$). The inviscid models can be run stably for all three parameterizations for a range of scaling coefficients $\gamma$. However, the optimal SSH metric is achieved when the ZB20-based parameterizations are turned off, i.e., $\gamma=0$ when $C_S=0$. This demonstrates that the ZB20-based parameterizations describe only part of the subgrid forcing, and cannot be used without an eddy viscosity model. Another important observation -- the optimal scaling coefficient $\gamma$ should be increased when the eddy viscosity coefficient $C_S$ is increased. Finally, the SSH bias can be efficiently reduced by the ZB20-Smooth and ZB20-Reynolds parameterizations for various values of the Smagorinsky coefficient ($C_S=0.03, 0.06, 0.09$).

\section*{Open Research}
The original NW2 configuration is available via \citeA{NW2code} and the data for the NW2 reference simulations, see \citeA{NW2data}. The version of the MOM6 source code with the implemented ZB20 parameterization, the configuration files for Double Gyre and NW2, and functions needed for generating the figures in this manuscript are available via \citeA{Perezhogin_code}. The simulation data is available via \citeA{Perezhogin_data}.

\acknowledgments

This project is supported by Schmidt Sciences, LLC.
C.F.G. was partially supported by NSF DMS Grant 2009752. This research was also supported in part through the NYU IT High Performance Computing resources, services, and staff expertise and by the National Science Foundation under Grant No. NSF PHY-1748958.
The authors would like to thank the members of M$^2$LInES, especially Dhruv Balwada and Nora Loose, for their helpful comments and discussions; Elizabeth Yankovsky for help with setting up the backscatter parameterization and NW2 simulations; Robert Hallberg and Marshall Ward for help with the MOM6 code. We also thank the editor, Tapio Schneider, as well as two anonymous reviewers and Julian Mak for their constructive comments that helped to improve the quality and presentation of this paper.



%
\bibliography{output}
%




%
%
%
%
%

\end{document}